\documentclass[twocolumn,prapplied,amsmath,amssymb,superscriptaddress]{revtex4-1}

\usepackage{graphicx}
\graphicspath{
	{.}
}

\usepackage{upgreek} 
\usepackage[caption=false]{subfig}
\usepackage{hyperref}
\usepackage{cleveref}
\usepackage{physics}
\usepackage{float}

\newcommand{\comsol}{COMSOL Multiphysics\textsuperscript{\textregistered}}
\newcommand{\hc}{\mathrm{h.c}}

\usepackage{xcolor}
\usepackage[todonotes={textsize=small},defaultcolor=cyan,authormarkuptext=name,commentmarkup=todo,final]{changes}
\definechangesauthor{0}
\definechangesauthor{1}
\definechangesauthor{2}
\definechangesauthor{3}
\definechangesauthor{4}
\definechangesauthor{5}
\definechangesauthor{6}
\definechangesauthor{7}
\definechangesauthor[name={}]{typo}

\let\oldcite\cite
\renewcommand{\cite}[1]{\mbox{\oldcite{#1}}}

\setdeletedmarkup{}

\newcommand{\phantomsubfloat}[1]{
	{
		\captionsetup[subfigure]{labelformat=empty}
		\subfloat[][]{#1}
	}%
}

\begin{document}
    \title{Level attraction from interference in two-tone driving}

    \author{Alan Gardin}
    \email{alan.gardin@adelaide.edu.au}
    \affiliation{%
    	Quantum and Nano Technology Group, School of Chemical Engineering, The University of Adelaide, Adelaide SA 5005, Australia
    }%
    \affiliation{%
    	IMT Atlantique, Lab-STICC, UMR CNRS 6285, F-29238 Brest, France
    }%
	\author{Guillaume Bourcin}
	\affiliation{%
		IMT Atlantique, Lab-STICC, UMR CNRS 6285, F-29238 Brest, France
	}%
	\author{Christian Person}
	\affiliation{%
		IMT Atlantique, Lab-STICC, UMR CNRS 6285, F-29238 Brest, France
	}%
	\author{Christophe Fumeaux}
	\affiliation{%
		School of Electrical Engineering and Computer Science, The University of Queensland, Brisbane QLD 4072, Australia
	}%
	\author{Romain Lebrun}
	\affiliation{%
		Unité Mixte de Physique, CNRS, Thales, Université Paris-Saclay, 91767 Palaiseau, France
	}%
	\author{Isabella Boventer}
	\affiliation{%
		Unité Mixte de Physique, CNRS, Thales, Université Paris-Saclay, 91767 Palaiseau, France
	}%
    \author{Giuseppe C. Tettamanzi}
	\affiliation{%
		Quantum and Nano Technology Group, School of Chemical Engineering, The University of Adelaide, Adelaide SA 5005, Australia
	}%
	\author{Vincent Castel}
	\affiliation{%
		IMT Atlantique, Lab-STICC, UMR CNRS 6285, F-29238 Brest, France
	}%

    \begin{abstract}
		Coherent and dissipative couplings, respectively characterised by energy level repulsion and attraction, each have different applications for quantum information processing. Thus, a system in which both coherent and dissipative couplings are tunable on-demand and in-situ is tantalising. A first step towards this goal is the two-tone driving of two bosonic modes, whose experimental signature was shown to exhibit controllable level repulsion and attraction by changing the phase and amplitude of one drive. However, whether the underlying physics is that of coherent and dissipative couplings has not been clarified, and cannot be concluded solely from the measured resonances (or anti-resonances) of the system. Here, we show how the physics at play can be analysed theoretically. Combining this theory with realistic finite-element simulations, we deduce that the observation of level attraction originates from interferences due to the measurement setup, and not dissipative coupling. Beyond the clarification of the origin of level attraction attributed to interference, our work demonstrates how effective Hamiltonians should be derived to appropriately describe the physics. 
    \end{abstract}
    \maketitle

	\section{Introduction}
	The coherent coupling between two systems corresponds to the coherent exchange of energy between them, and is characterised by energy level repulsion (also known as normal-mode splitting or an anti-crossing). This phenomenon is ubiquitous in physics, spanning the classical coupling of two harmonic oscillators \cite{2010Novotny} to the coupling of bosonic quasi-particles with two-level systems \cite{1992Weisbuch} or other bosonic modes \cite{2008Dobrindt}. With respect to quantum information processing, coherent coupling allows converting quantum information between light and solid-state degrees of freedom, and is therefore an elementary building block of quantum communication \cite{2007Gisin,2008Kimble,2019LachanceQuirion}. On the other hand, dissipative coupling \cite{2023Lu,2020Wang,2021HarderHu}, characterised by energy level attraction instead, arises due to the indirect coupling of two modes mediated by a common reservoir \cite{2019Wang,2015Metelmann,2022Clerk} (e.g. a strongly dissipative auxiliary mode \cite{2019YuWangYuanXiao}, a photonic environment \cite{2019YaoCommPhys,2019Yao,2024Bleu}, metallic leads \cite{2002Kubala}). The merging of energy levels characterising dissipative couplings leads to exceptional points \cite{2012Heiss,2017Zhang,2019ZhangYou,2022HurstFlebus}, which can be useful for topological energy transfer \cite{2016Xu}, and improved sensitivity for quantum metrology and quantum sensing applications \cite{2019CaoYanPRB,2020Yu}. Furthermore, balancing coherent and dissipative couplings, using dissipation engineering and synthetic gauge fields for instance, allows breaking time-reversal symmetry and thus create non-reciprocal devices \cite{2019Wang,2015Metelmann,2022Clerk,2010Koch,2015Sliwa,2017Fang,2021Huang}. Therefore, building a system in which both coherent and dissipative couplings co-exist and are tunable would be useful to access all the aforementioned applications in a unique versatile platform. \added[id=1]{However, while both coherent and dissipative couplings can co-exist in a single platform, their tunability is usually not practical. For instance, in cavity magnonics implementations \cite{2019Yao,2019YaoCommPhys}, this requires mechanical rotation of the cavity or of a static magnetic field.}
	
	\replaced[id=1]{In a system with both coherent and dissipative couplings, the energy spectrum should exhibit both level repulsion and attraction, respectively}{Such a system, due to the presence of both coherent and dissipative couplings, should have an energy spectrum exhibiting both level repulsion and level attraction}. A two-tone driving scheme proposed by \citeauthor{2018Grigoryan} \cite{2018Grigoryan} suggested theoretically that such an energy spectrum could be obtained by driving simultaneously two coherently-coupled bosonic modes. In a subsequent work, the same authors predicted a tunable cavity-mediated dissipative coupling taking advantage of the same principle \cite{2019Grigoryan}.
	Furthermore, two experiments by \citeauthor{2019Boventer} \cite{2019Boventer,2020BoventerPRR} implemented the original proposal \cite{2018Grigoryan}, and confirmed tunable level repulsion and attraction in the reflection spectrum.
	
	These results were promising, since it is usually expected that the response of a system to excitations at frequencies close to their normal modes leads to resonances (or anti-resonances), with various possible line shapes informing on the underlying physics \cite{2010Baernthaler,2017Limonov}. \added[id=1]{Notably, in these experiments \cite{2019Boventer,2020BoventerPRR}, the control is performed through the phase and amplitude of a microwave drive, which is better for integration, for instance compared to the vector magnet required in \cite{2019Yao,2019YaoCommPhys}.}
	However, inspection of the experimental signature alone is not sufficient because it may not be related to the energy levels.
	For instance, while dissipative coupling implies level attraction in a variety of platforms such as optomechanics \cite{2023Lua}, Aharonov-Bohm interferometers \cite{2002Kubala} or semiconductor microcavities \cite{2024Bleu}, alternative physics can also lead to level attraction, as exemplified in optomechanics \cite{2018Bernier} or spinor condensates \cite{2014Bernier}. 
	Furthermore, the model used in the original proposal \cite{2018Grigoryan} was phenomenological, and as such it only reproduces the experimental observable, but cannot be used to conclude on the underlying physics.
	Thus, the physics at play in two-tone driving remains unclear. 
	
	
	In this work, we model a two-tone driving experiment, similar to those of \cite{2019Boventer,2020BoventerPRR}. Our starting point is the \added[id=2]{closed-system} Hamiltonian of two coherently coupled bosonic modes, whose spectrum is that of level repulsion. To model the two-tone driving and the experimentally accessible quantities, we use quantum Langevin equations (QLEs) and the input-output formalism \cite{1985GardinerCollett} (\cref{sec:model}). This theoretical treatment allows to find an analytical expression for the experimental signature, where both level repulsion and attraction can occur\added[id=2]{, in contrast with the sole level repulsion expected from the closed-system Hamiltonian}. In \cref{sec:analytical}, we use the analytical expressions of \cref{sec:model} to understand the origin of the tunable level repulsion and attraction. These insights guide our finite-element simulations of a realistic setup, and allows us to confirm that level attraction can be attributed to an anti-resonance due to the destructive interference between the reflection and transmission coefficients. \added[id=2]{While this explains the level attraction, it does not completely eliminate the possibility of dissipative coupling.} To clarify the physics \added[id=2]{in the presence of the drives}, we derive the open-system effective Hamiltonian of the two-tone driven system, and show that the physics remains exclusively that of coherent coupling, despite the observation of level attraction. We conclude in \cref{sec:conclusion}.

	\begin{figure}[t]
		\centering
		\includegraphics[width=\linewidth]{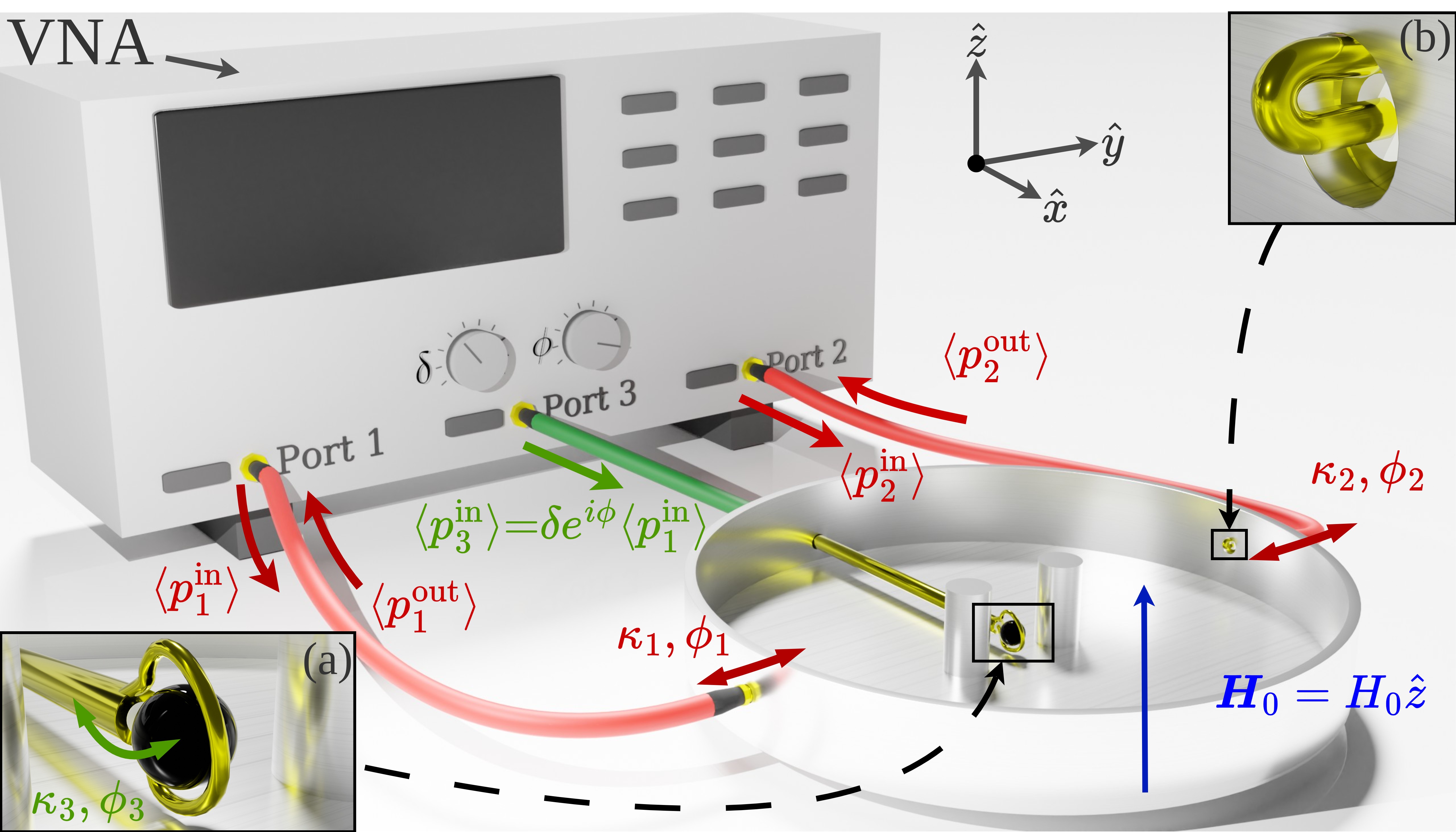}
		\caption{Schematic of a two-tone driving experiment using a two-post re-entrant cavity \cite{2020Bourhill} loaded with a ferromagnetic sphere made of Yttrium-Iron-Garnet. Ports 1 and 3 are always active and drive respectively the cavity and magnon modes. The cavity is connected to a vector network analyser (VNA) to measure the reflection at Port 1 and the transmission through Port 2. Inset (a): the YIG sphere is placed at the centre of a loop made by soldering the inner and outer conductor of a coaxial cable, itself placed between the two posts where the cavity mode's magnetic field is strongest. In the picture, the loop is in the $(x,z)$ plane, and the YIG sphere is biased along $\vu{z}$ by the static magnetic field $\vb{H}_0$. Inset (b): antenna used to excite the cavity modes.}
		\label{fig:experiment-schematic}
	\end{figure}

	\section{\label{sec:model}Theoretical model and experimental observables}
	
	\paragraph{Quantum Langevin equations.}
	We consider the experimental setup of \cref{fig:experiment-schematic}, where a ferromagnetic sphere made of Yttrium-Iron-Garnet (YIG) is placed inside a microwave cavity. The second-quantised Hamiltonian of the system enclosed inside the cavity is that of the lowest-order magnon mode (quantised spin wave) coupled to a single cavity mode, described by \cite{2021HarderHu}
	\begin{equation}
		\label{eq:system-hamiltonian}
		H_\text{sys} = \hbar \omega_c c^\dagger c + \hbar \omega_m m^\dagger m
		+ \hbar \qty(g cm^\dagger + g^* c^\dagger m),
	\end{equation}
	where $g/2\pi$ is the (complex) coherent coupling strength \cite{2019Flower,2023Gardin} between the cavity mode (operators $c,c^\dagger$) and the magnon mode (operators $m,m^\dagger$). The frequency of the cavity mode $\omega_c/2\pi$ is fixed by the cavity's geometry, while the ferromagnetic resonance of the magnon is tuned by a static magnetic field $\vb{H}_0 = H_0 \vu{z}$ as $\omega_m = \gamma \abs{\vb{H}_0}$ where ${\gamma}{/2\pi} = 28$ GHz/T is the gyromagnetic ratio. 
	
	The quantum Langevin equations (QLEs) allow to model the coupling of the cavity system described by \cref{eq:system-hamiltonian} to the environment, to model intrinsic dissipation or external drives. 
	To model the intrinsic damping of the cavity mode $c$ and the magnon mode $m$, we assume that they each couple to a different bosonic bath with coupling constants $\kappa_c/2\pi$ and $\kappa_m/2\pi$. 
	As per \cref{fig:experiment-schematic}, the cavity system described by \cref{eq:system-hamiltonian} is coupled to three ports $p_i$, labelled from 1 to 3, where the first two ports couple only to the cavity mode (with real-valued coupling constants $\kappa_1/2\pi$ and $\kappa_2/2\pi$ and phases $\phi_1,\phi_2$ \cite{2024Bourcin}) and the third couples only to the magnon mode (real-valued coupling constant $\kappa_3/2\pi$ with phase $\phi_3$).
	Physically, the coupling of the magnon to port 3 is due to the Zeeman interaction with the magnetic field created by the loop of port 3. 
	The QLE for the cavity and magnon modes then read (see \cref{app:qle} for details)
	\begin{align}
		\label{eq:qle-c}
		\dot{c}
		&= -i \widetilde \omega_c c -ig^* m
		- \sqrt{\kappa_c}c^\text{in}
		- \sqrt{\kappa_1} e^{i \phi_1} p_1^\text{in}
		- \sqrt{\kappa_2} e^{i \phi_2} p_2^\text{in},\\
		\label{eq:qle-m}
		\dot{m}
		&= -i \widetilde \omega_m m -ig c
		- \sqrt{\kappa_m}m^\text{in}
		- \sqrt{\kappa_3} e^{i \phi_3} p_3^\text{in},
	\end{align}
	where $\widetilde{\omega}_c = \omega_c-i\frac{\kappa_c + \kappa_1+\kappa_2}{2}$ and $\widetilde{\omega}_m = \omega_m-i\frac{\kappa_m + \kappa_3}{2}$. In the QLEs, $c^\text{in}$ and $m^\text{in}$ account for intrinsic damping and have zero mean $\expval {c^\text{in}}=\expval {m^\text{in}}=0$, while $p_i^\text{in}$ represent the inputs from the ports.

	\paragraph{Two-tone reflection and transmission.}
	We now assume that $p_1^\text{in}$ and $p_3^\text{in}$ correspond to coherent drives at the same frequency, albeit with a phase and amplitude difference written $\expval{p_3^\text{in}} = \delta e^{i\phi} \expval{p_1^\text{in}}$. We also perform a semi-classical approximation and neglect quantum fluctuations, which amounts to only considering expectation values.
	Using the input-output formalism (see \cref{app:input-output} for details), the reflection at Port 1 when Port 3 is active is found to be
	\begin{align}
		r_{1}(\omega) 
		&= 
		\frac{
			\left. 
			\expval{ p_1^\text{out}}
			\right|_{\expval{p_2^\text{in}}=0}
		}{\expval{ p_1^\text{in}}}\\
		\label{eq:input-output-simple}
		&= 1 
		-i \sqrt{\kappa_1}\frac{
			\widetilde{\Delta}_m \sqrt{\kappa_1}
			+ g^*\sqrt{\kappa_3} e^{i(\phi_3- \phi_1)} \delta e^{i \phi} 
		}{\widetilde{\Delta}_c \widetilde{\Delta}_m - \abs{g}^2}
	\end{align}
	where $\widetilde{\Delta}_c = \omega-\widetilde\omega_c$ and $\widetilde{\Delta}_m = \omega-\widetilde\omega_m$.
	By comparing with the expressions of the standard S-parameters (detailed in \cref{app:R1}), we note that the reflection $r_1$ can also be written $r_1(\omega) = S_{11} + \delta e^{i \phi} S_{13}$
	which is expected given the linearity of the problem. 

	\begin{figure}[t]
		\centering
		\includegraphics[width=\linewidth]{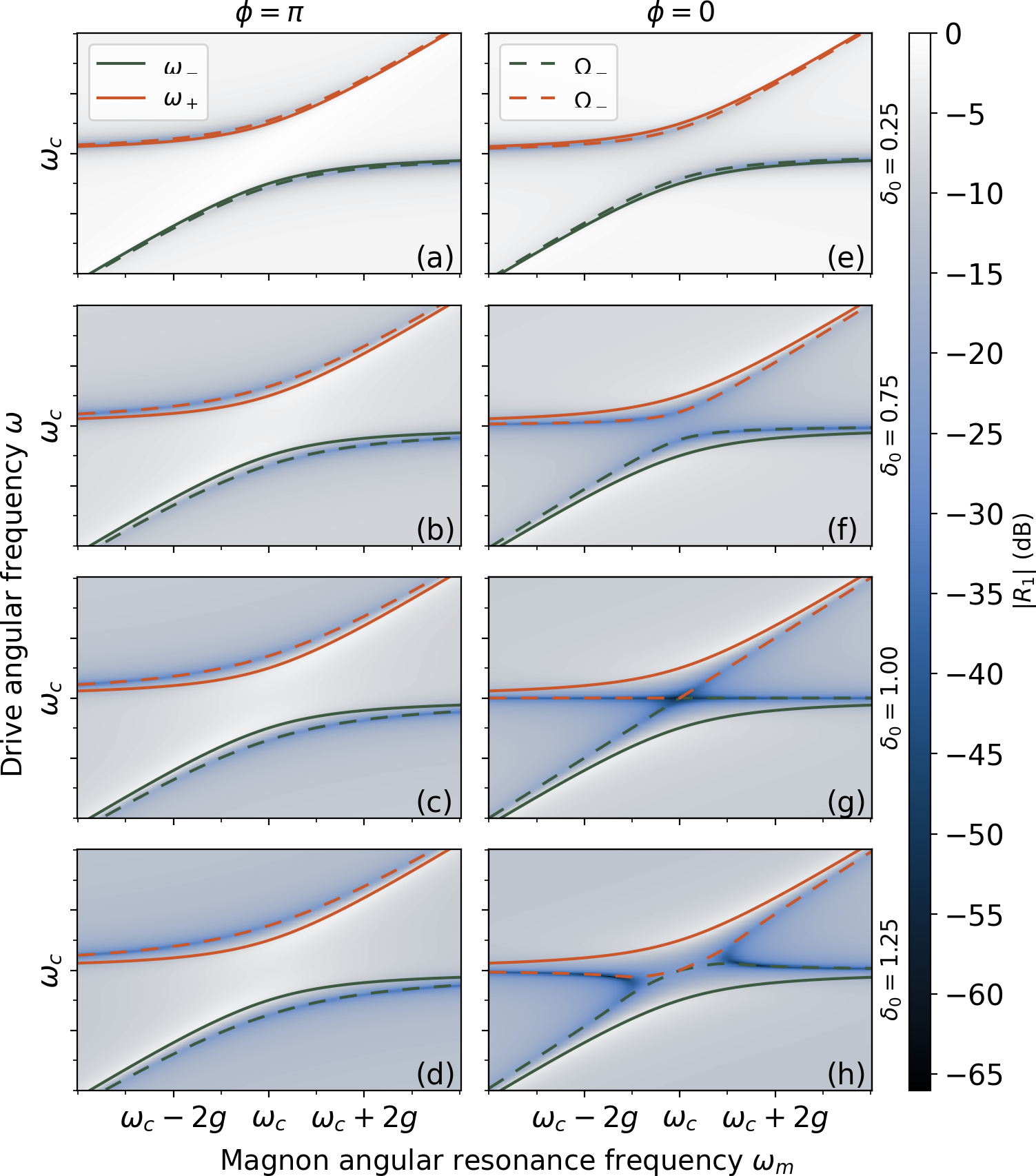}
		\phantomsubfloat{\label{fig:reflection:a}}
		\phantomsubfloat{\label{fig:reflection:b}}
		\phantomsubfloat{\label{fig:reflection:c}}
		\phantomsubfloat{\label{fig:reflection:d}}
		\phantomsubfloat{\label{fig:reflection:e}}
		\phantomsubfloat{\label{fig:reflection:f}}
		\phantomsubfloat{\label{fig:reflection:g}}
		\vspace{-2.5em}
		\caption{Normalised reflection amplitude $\abs{R_{1}}$ at Port 1 for increasing values of $\delta_0 = \frac{\sqrt{\kappa_1 \kappa_3}}{\abs{g}} \delta$ from top to bottom, and a dephasing $\phi=\pi$ (first column) and $\phi=0$ (second column) between the drives of Port 1 and Port 3. The solid lines correspond to the real part of the zeros $\widetilde \omega_\pm$ of the denominator of $R_{1}$, given by \cref{eq:zeros-denom-s11}, while the dashed lines correspond to the zeros $\widetilde \Omega_\pm$ of the nominator of $R_1$. }
		\label{fig:reflection}	
	\end{figure}
	
	For $\delta \neq 0$, the reflection coefficient $r_1$ 
	can be greater than one, because it is only normalised to the input power from Port 1, thus neglecting that of Port 3. Following the setup of \cref{fig:experiment-schematic}, and taking the power out of Port 1 as a reference, the VNA outputs a power $1+\delta^2$, so we can renormalise the reflection to $R_{1}(\omega) = \frac{1}{\sqrt{1+\delta^2}} r_{1}(\omega)$. 
	Similarly, we can calculate the normalised transmission coefficient through Port 2 when both Port 1 and Port 3 are active, and we find (\cref{app:transmission})
	\begin{align}
		T_{2}(\omega) 
		&= 
		\frac{1}{\sqrt{1+\delta^2}}
		\frac{
			\left. 
			\expval{p_2^\text{out}}
			\right|_{\expval{p_2^\text{in}}=0}
		}{\expval{ p_1^\text{in}}}
		= \frac{S_{21} + \delta e^{i\phi} S_{23}}{\sqrt{1+\delta^2}}.
	\end{align}
	
\begin{figure}[t]
		\centering
		\includegraphics[width=\linewidth]{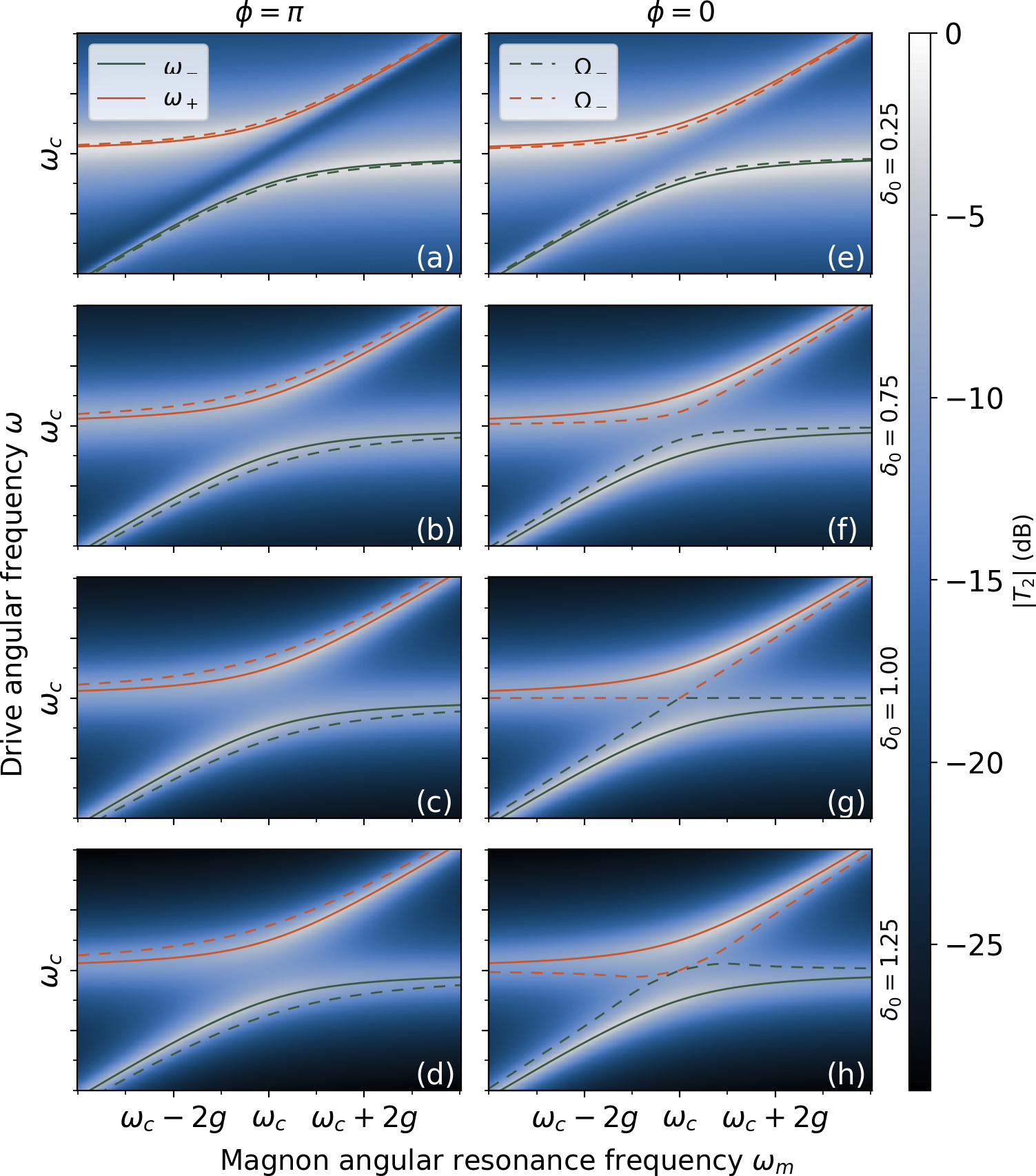}
		\caption{Normalised transmission amplitude $\abs{T_2}$ through Port 2 for $\phi=\pi$ and $\phi=0$, and increasing values of $\delta_0$ from top to bottom. The parameters \added[id=3]{and legends} used are identical to those of \cref{fig:reflection}.}
		\label{fig:transmission}	
	\end{figure}

	We plot $\abs{R_1}$ in \cref{fig:reflection} and $\abs{T_2}$ in \cref{fig:transmission} with the parameters $\omega_c/2\pi=10$ GHz, $\abs{g}/2\pi=20$ MHz, $\kappa_c/2\pi=\kappa_m/2\pi=1$ MHz, $\kappa_1/2\pi= \kappa_2/2\pi=10$ MHz, $\kappa_3/2\pi=5$ MHz, \replaced[id=4]{$\varphi=\mathrm{arg}\, g = -\frac{\pi}{2}$}{$\varphi = -\pi/2$} and $\phi_1=\phi_2=\phi_3=0$. The reflection $R_1$ exhibits controllable level repulsion and attraction depending on the amplitude $\delta$ of the drives and the dephasing $\phi$, mirroring the experimental results of \cite{2019Boventer,2020BoventerPRR}. On the other hand, we notice that the transmission $T_2$ only shows level repulsion, despite the system being driven in exactly the same way: between $R_1$ and $T_2$, only the measurement location changes, and thus the physics should be the same. \added[id=2]{In the next section, we will attribute the observed level attraction to interference.}

	\section{\label{sec:analytical}Analysis and validation by finite-element simulations}
	\paragraph{Analysis of the two-tone reflection coefficient.} An advantage of the input-output formalism is that it allows us to understand analytically the spectral features of \cref{fig:reflection}. Indeed, by writing $r_1(\omega) = N(\omega)/D(\omega)$ in terms of a nominator $N$ and a denominator $D$,
	the denominator can be factorised as
	\begin{align}
		D(\omega)
		\label{eq:quadratic-denom}
		&=
		\omega^2 
		- \qty(\widetilde \omega_c+ \widetilde \omega_m) \omega
		+ \widetilde \omega_c \widetilde \omega_m - \abs{g}^2\\
		&= 
		\qty(\omega - \widetilde \omega_+)
		\qty(\omega - \widetilde \omega_-)
	\end{align}
	where
	\begin{equation}
		\label{eq:zeros-denom-s11}
		\widetilde \omega_\pm = 
		\frac{\widetilde \omega_c + \widetilde \omega_m}{2}
		\pm \frac{\sqrt{
				\qty(\widetilde \omega_c- \widetilde \omega_m)^2
				+ 4\abs{g}^2
		}}{2}
	\end{equation}
	are the complex-valued solutions of the quadratic \cref{eq:quadratic-denom}. 
	
	On the other hand, the nominator $N$ can be written as
	\begin{align}
		N(\omega)
		\label{eq:quadratic-nom}
		&=
		\omega^2
		- \qty(\widetilde \omega_c' + \widetilde \omega_m) 
		\omega
		+ 
		\widetilde \omega_c' \widetilde \omega_m
		-G^2
	\end{align}
	where we defined $\widetilde \omega_c' = \widetilde \omega_c + i \kappa_1$ and
	\begin{align}
		G 
		\label{eq:effective-coupling-strength}
		&=
		\abs{g} \sqrt{1 + \frac{\sqrt{\kappa_1 \kappa_3}}{\abs{g}} \delta e^{i\qty(\phi + \phi_3-\phi_1 + \frac{\pi}{2}- \arg{g})}},
	\end{align}
	with $\arg{g}$ is the phase of the complex number $g$.
	Comparing \cref{eq:quadratic-denom,eq:quadratic-nom}, we see that the nominator can be factored similarly to the denominator as $N(\omega)=(\omega-\widetilde \Omega_+(\delta, \phi))(\omega- \widetilde\Omega_-(\delta, \phi))$, where the complex frequencies $\widetilde \Omega_\pm(\delta, \phi)$ are formally identical to \cref{eq:zeros-denom-s11}, after replacing $\omega_c \mapsto \omega_c'$ and $\abs{g} \mapsto G$. Finally, we obtain
	\begin{equation}
		\label{eq:r1:factored}
		r_{1}(\omega)
		= 
		\frac{
			\qty(\omega-\widetilde \Omega_+(\delta, \phi))\qty(\omega-\widetilde \Omega_-(\delta, \phi))
		}{
			\qty(\omega-\widetilde \omega_+)
			\qty(\omega- \widetilde \omega_-)
		}.
	\end{equation}

	In general, the observed resonances and anti-resonances of \cref{eq:r1:factored} are a combined effect of both the numerator $N$ and the denominator $D$. However, as discussed below, and numerically illustrated in \cref{fig:reflection}, the spectral features are well characterised by the poles (zeroes $\widetilde{\omega}_\pm$ of $D$) and zeroes (zeroes $\widetilde{\Omega}_\pm$ of $N$) of $r_{1}$. Notably, it is sufficient to solely examine the \replaced[id=5]{zeroes}{poles} $\widetilde{\Omega }_{\pm}$ to understand the resonances, and the \replaced[id=5]{poles}{zeroes} $\widetilde{\omega }_{\pm}$ to understand the anti-resonances, independently of each other.

	For small dissipation rates $\kappa_i$, the imaginary part of $\widetilde \omega_\pm$ is small compared to its real part. Therefore, in \cref{eq:r1:factored}, when $\omega$ is close to $\omega_\pm = \Re{\widetilde \omega_\pm}$, the denominator $\abs{D}$ of $\abs{R_1}$ almost vanishes, leading to a resonance behaviour (maxima) of $\abs {R_1}$. Furthermore, $\omega_\pm$ corresponds to the spectrum of the Hamiltonian of \cref{eq:system-hamiltonian}, and hence this resonance behaviour is expected to give information about the spectrum of the closed-system. As shown by the solid lines in \cref{fig:reflection,fig:transmission}, the spectrum is that of coherent coupling, characterised by energy level repulsion with an angular frequency gap $2\abs{g}$. Hence, the denominator of $\abs{R_1}$, or equivalently its resonances, does inform on the underlying physics.


	The situation is different for the nominator $N$ of $R_1$, which leads to anti-resonances. 
	Indeed, while the expressions of $\widetilde \omega_\pm$ and $\widetilde \Omega_\pm$ are formally identical, the anti-resonance coupling strength $G/2\pi$ for $\widetilde \Omega_\pm$ is complex-valued (while it is real-valued, $\abs{g}/2\pi$, for $\widetilde \omega_\pm$) which can lead to level attraction. To see this more clearly, it is convenient to introduce the effective amplitude $\delta_0 =\frac{\sqrt{\kappa_1 \kappa_3}}{\abs{g}} \delta$ and the effective phase $\phi_0 = \phi_1-\phi_3 + \arg{g}-\frac{\pi}{2}$ of Port 3. We can then rewrite \cref{eq:effective-coupling-strength} as
	\begin{equation}
	\label{eq:effective-coupling-strength-delta0}
		G =g \sqrt{1 + \delta_0 e^{i\qty(\phi- \phi_0)}}.
	\end{equation}
	Therefore, when $\phi-\phi_0=0$ the anti-resonance coupling strength $G/2\pi$ is real-valued and increases as $\delta_0$ increases, leading to an increase in level repulsion (see the dashed lines in the first column of \cref{fig:reflection}). On the other hand, when $\phi-\phi_0=\pi$, $G/2\pi$ is real-valued when $\delta_0<1$ and diminishes when $\delta_0$ increases. Eventually, when $\delta_0>1$, the $G/2\pi$ becomes purely imaginary due to taking the square root of a negative number, leading to level attraction. In the limiting case where $\delta_0=1$, we have $G=0$ and we obtain two uncoupled anti-resonances as shown in \cref{fig:reflection:g} (horizontal at the cavity mode frequency and diagonal at the magnon's frequency $\omega_m/2\pi$).
	As can be seen from the dashed lines in \cref{fig:reflection}, these spectral features indeed correspond to anti-resonances (
	minima of $\abs{R_1}$) at the frequencies $\Omega_\pm$. Physically, these frequencies $\Omega_\pm$ are determined by the interference between the nominators of $S_{11}$ and $S_{13}$ since $r_1(\omega) = S_{11} + \delta e^{i \phi} S_{13}$. Hence, the input-output formalism shows that while the denominator of $\abs{R_1}$ informs on the physics, the nominator is interference-based. Furthermore, the denominators of both $R_1$ and $T_2$ are identical (see \cref{app:transmission} for the detailed expression), leading to similar resonant behaviour following the coherent coupling spectrum. However, their anti-resonance behaviours differ because their nominators differ (see \cref{app:transmission}).
	
	\begin{figure}[t]
		\centering
		\includegraphics[width=\linewidth]{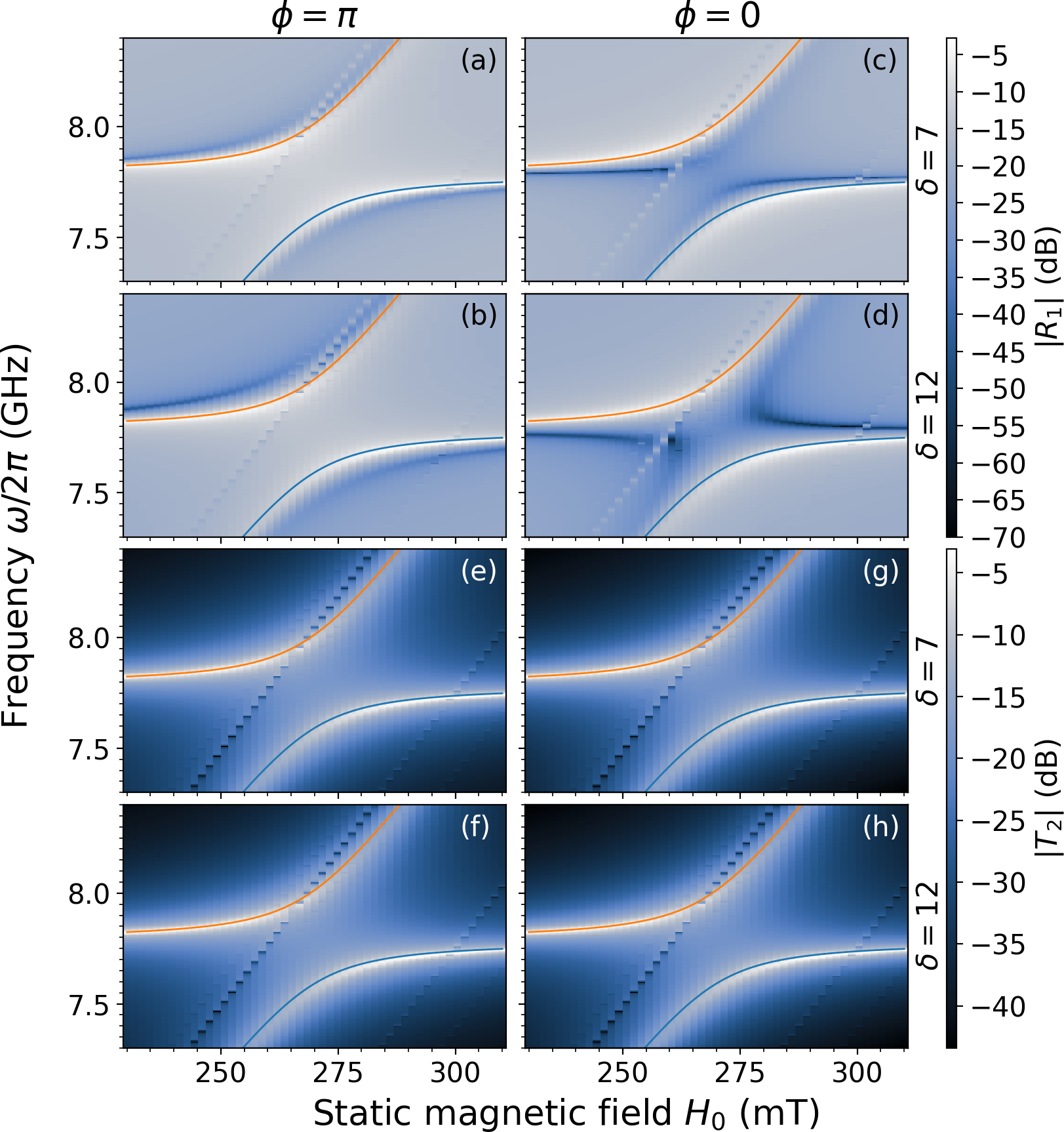}
		\caption{Plots of the normalised reflection and transmission amplitudes $\abs{R_1},\abs{T_2}$ using the numerical values of the S parameters obtained using COMSOL. Note that contrary to \cref{fig:reflection} we varied the amplitude $\delta$ at Port 3 instead of the effective amplitude $\delta_0$, because the latter depends on the unknown parameters $\kappa_1,\kappa_3$.}
		\label{fig:comsol}	
		\vspace{-1em}
	\end{figure}

	Concerning the occurrence of level attraction, we note that it requires the total phase $\phi+\phi_3-\phi_1+\frac{\pi}{2} - \mathrm{arg} \, g$ equal to $\pi$. Physically, this is determined by the magnon-photon coupling phase $\mathrm{arg} \, g$, the coupling of the cavity and magnon modes to the probes ($\phi_1$ and $\phi_3$), and the chosen \replaced[id=0]{phase difference}{dephasing} $\phi$ between Port 1 and Port 3. The strength of level attraction itself is determined by $|G|$, which depends on $|g|$ and the ratio $\frac{\sqrt{\kappa_1 \kappa_3}}{g}$, i.e. the magnon-photon coupling strength and the coupling to the ports.

	\paragraph{Numerical finite element results.}
	The input-output formalism provides interesting insights thanks to the resulting analytical expressions. However, it remains a toy-model of a two-tone driving experiment. Instead, a realistic modelling, taking into account the geometry of the cavity and of the ports, can be perfomed using \comsol, a finite element modelling software. We use a two-post re-entrant cavity \cite{2014Goryachev,2016Kostylev}, similar to that sketched in \cref{fig:experiment-schematic}, in which the loop antenna of Port 3 is inserted from the bottom of the cavity (see \cref{app:comsol} for more details). At the location of the YIG sphere, the cavity mode's magnetic field is purely along the $\vu{x}$ axis \cite{2020Bourhill}. In our analysis, we assumed that Port 3 does not couple to the cavity mode, which requires us to orient the loop antenna so that it generates a magnetic field orthogonal to the cavity mode.  Thus, noting $\vu{n}$ the normal to the plane of the loop, we need $\vu n$ to be in the $(x,z)$ plane. Furthermore, since only the components orthogonal to the static magnetic field $H_0 \vu{z}$ contribute to the coupling between Port 3 and the magnon, $\kappa_3/2\pi$ is maximum when $\vu n = \vu y$. 
	
	We first numerically calculated the S parameters for different values of $\vb H_0$ using COMSOL, as detailed in \cref{app:comsol}. From these two $S$ parameters, we can plot $R_1$ and $T_2$ after normalising $r_1(\omega) = S_{11} + \delta e^{i \phi} S_{13}$ and $t_2(\omega) = S_{21} + \delta e^{i \phi} S_{23}$, and we obtain the results of \cref{fig:comsol}, which successfully reproduce the input-output theory results of \cref{fig:reflection,fig:transmission}. As a further verification, we also performed a fully two-tone experiment in COMSOL for $\delta=12$, and the results were identical to those of the second and fourth rows of \cref{fig:comsol}.

	\section{\label{sec:physics}Physical interpretation}
	In \cref{sec:model,sec:analytical}, we have considered the experimentally accessible observables, i.e. the reflection and transmission amplitudes.
	We have confirmed that the resonances of $R_1$ and $T_2$ follow energy level repulsion, suggesting coherent coupling physics. As the same time, we saw that the anti-resonances of $R_1$ can lead to level attraction, which begs the question of which physics is effectively taking place. The physics can be inferred from the dynamics of the cavity and magnon modes in the presence of the two coherent drives, given by the quantum Langevin equations.
	Therefore, to derive the effective Hamiltonian, we write the QLEs \cref{eq:qle-c,eq:qle-m} in the semi-classical approximation where operators are replaced by their expectation values. Recalling that $\expval {c^\text{in}} = \expval {m^\text{in}}=0$, the QLEs read
	\begin{align}
		\dot{c}
		&= -i \widetilde{\omega}_c  c -ig m
		- \sqrt{\kappa_1} e^{i \phi_1} p_1^\text{in}
		- \sqrt{\kappa_2} e^{i \phi_2} p_2^\text{in},\\
		\dot{m}
		&= -i \widetilde{\omega}_m m -ig c
		- \sqrt{\kappa_3} e^{i \phi_3} p_3^\text{in}.
	\end{align}
	In both a reflection and transmission measurement, port 2 is not driven, so we set $p_2^\text{in}=0$. On the other hand, $p_1^\text{in}$ and $p_3^\text{in}$ correspond to coherent drives at frequency $\omega/2\pi$ with amplitudes $\mathcal E$ and $\delta e^{i\phi}\mathcal E$. Thus, further defining $\mathcal E_1 = \sqrt{\kappa_1/2\pi}e^{i \phi_1} \mathcal E$ and $\mathcal E_3 = \sqrt{\kappa_3/2\pi}e^{i \phi_3} \mathcal E \delta e^{i \phi}$, the QLEs become, from \cref{app:input-output},
	\begin{align}
		\label{eq:effective-c}
		\dot{c}
		&= -i \widetilde{\omega}_c  c -ig m
		- \mathcal E_1 e^{-i\omega t},\\
		\label{eq:effective-m}
		\dot{m}
		&= -i \widetilde{\omega}_m m -ig c
		- \mathcal E_3 e^{-i\omega t}.
	\end{align}
	From these equations, we find that the effective non-hermitian Hamiltonian (recall that $\widetilde \omega_c, \widetilde \omega_m \in \mathbb C$)
	\begin{equation}
		\label{eq:effective-hamiltonian}
		\begin{aligned}
			H_\text{eff}
			&=
			\hbar \widetilde \omega_c c^\dagger c 
			+ \hbar \widetilde \omega_m m^\dagger m
			+ \hbar \qty(g cm^\dagger + g^* c^\dagger m)\\
			\quad&
			+i\hbar \qty(\mathcal E_1 c e^{i\omega t} - \hc)
			+i\hbar \qty(\mathcal E_3 m e^{i\omega t} - \hc)
		\end{aligned}
	\end{equation}
	where $\hc$ stands for the hermitian conjugate terms, gives Heisenberg equations of motion identical to \cref{eq:effective-c,eq:effective-m}. Thus, the physics described by \cref{eq:effective-hamiltonian} is identical to that given by the QLEs. After a rotating frame transformation to remove the time-dependence and displacement operations (see \cref{app:two-tone-coherent-drive} for details), this Hamiltonian is unitarily equivalent to
	\begin{equation}
		H_\text{eff}'
		=
		-\hbar \widetilde \Delta_c c^\dagger c 
		- \hbar \widetilde \Delta_m m^\dagger m
		+ \hbar \qty(g cm^\dagger + g^* c^\dagger m).
	\end{equation}
	
	Formally similar to the closed-system Hamiltonian of \cref{eq:system-hamiltonian}, this effective Hamiltonian describes a coherent coupling physics, with a spectrum corresponding to level repulsion. Since $H_\text{eff}'$ describes the physics of the cavity system in the presence of coherent drives, we conclude that the physics of two-tone driving indeed corresponds to coherent coupling, even though the reflection coefficient can show level attraction due to interference-based anti-resonances. Therefore, the anti-resonance frequencies $\Omega_\pm/2\pi$ (dashed lines in \cref{fig:reflection}) do not correspond to the spectrum of the system, which are instead given by $\omega_\pm/2\pi$ (solid lines in \cref{fig:reflection}). Notably, the anti-resonance coupling $G/2\pi$ is merely a convenient quantity to understand the attraction and repulsion of the anti-resonances, but it does not represent a physical coupling.
	For instance, when $\delta_0=1$, the magnon and cavity modes are still coherently coupled with coupling strength $g/2\pi$, even though $G=0$ and level crossing is observed (see \cref{fig:reflection:g}).

	In the two-tone driving system studied here, the anti-resonances leading to level attraction come from interferences between $S_{11}$ and $S_{13}$, but anti-resonances can also appear in one-tone driven systems as experimentally demonstrated by \citeauthor{2019RaoYuHu} \cite{2019RaoYuHu}. Notably, these anti-resonances can exhibit different coupling behaviour recently analysed by \cite{2024Bourcin}, and suggest energy level repulsion and attraction with a magnon mode. These behaviours are associated with the numerators of the transmission coefficient, and therefore they are not associated to a physical coherent and dissipative coupling (even though they do allow to fit experimental data). Importantly, it is incorrect to derive an effective Hamiltonian from this numerator alone, as it is not associated with a physical dynamics.

	\section{\label{sec:conclusion}Conclusion}
	To conclude, we have showed that when two bosonic modes are simultaneously driven, the resonances in reflection and transmission indeed inform on the normal modes of the system, while the observed anti-resonances in reflection are due to interference physics, and are unrelated to the normal modes of the coupled system. \added[id=1]{Therefore, there is no dissipative coupling physics in this system, which makes the two-tone driving scheme unsuitable for the in-situ control of coherent and dissipative couplings.} \replaced[id=1]{Still, realising such an all-microwave control remains}{We note that while this two-tone driving does not correspond to the in-situ control of coherent and dissipative couplings, achieving it is still} a relevant research direction\added[id=1]{, due to the promising applications it would unlock}. 
	
	\added[id=1]{It is worth mentioning that an interference-based level attraction was recently used in a cavity magnonics system to experimentally achieve a nearly perfect single-beam absorption of 96 \% \cite{2021Rao}. However, it once again required a vector magnet, whereas in the present work, we achieved similar physics through a flexible two-tone driving instead. Thus, it is of interest to explore the possibilities offered by two-tone driving for coherent perfect absorption, which we leave for future work.}
	
	Finally, while we considered a cavity magnonics system as a physical realisation of two coupled bosonic modes, many other systems, such as intersubband polaritons \cite{2005Ciuti} or cavity optomechanics in the red-detuned regime \cite{2014Aspelmeyer}, reduce to a similar Hamiltonian (the Dicke model \cite{1973Hepp} after employing the Holstein-Primakoff transformation \cite{1940Holstein} and the rotating wave approximation \cite{2020LeBoite}). \replaced[id=0]{While the details of how such modes would be coherently driven would differ}{Hence}, the derivations employed here are very general and may prove useful to analyse the physics in other open systems.

	\begin{acknowledgments}
		We acknowledge financial support from Thales Australia and Thales Research and Technology. This work is part of the research program supported by the European Union through the European Regional Development Fund (ERDF), by the Ministry of Higher Education and Research, Brittany and Rennes Métropole, through the CPER SpaceTech DroneTech, by Brest Métropole, and the ANR projects ICARUS (ANR-22-CE24-0008) and MagFunc (ANR-20-CE91-0005). The scientific colour map \textit{oslo} \cite{2021Crameri} was used to prevent visual distortion of the data and exclusion of readers with colourvision deficiencies \cite{2020Crameri}.
	\end{acknowledgments}

	\appendix
	\onecolumngrid
	\section{\label{app:qle}Derivation of the quantum Langevin equations}
    \paragraph{System definition.}
    In this note, we derive the quantum Langevin equations of the coupled magnon/photon cavity system under the rotating wave approximation, which is valid for $\abs{g} \ll \omega_c,\omega_m$ \cite{2019FriskKockum,2020LeBoite}, where $g/2\pi$ is the coupling strength. We recall that the most general Hamiltonian describing a coupled magnon-photon system reads \cite{2019Flower} 
    \begin{equation}
    	H_\text{sys} = \hbar \omega_c c^\dagger c + \hbar \omega_m m^\dagger m
    	+ \hbar (gcm^\dagger + g^* c^\dagger m),
    \end{equation}
    where $g/2\pi$ is complex-valued due to potential coupling phases \cite{2023Gardin}.
    
    The intrinsic losses of the cavity and magnon modes (respectively due to radiative losses and the Gilbert damping) can be modelled by coupling to two bosonic baths described by the Hamiltonians
		\begin{align}
			H_\text{bath,intrinsic} 
			&=
			\int_\mathbb{R} \dd{\omega} \hbar \omega a_\omega^\dagger a_\omega
			+ 
			\int_\mathbb{R} \dd{\omega} \hbar \omega b_\omega^\dagger b_\omega,\\
			H_\text{cav--bath,intrinsic}
			&=
			\int_\mathbb{R} \dd{\omega} i\hbar \sqrt{\frac{\kappa_c}{2\pi}}
			\qty(ca_\omega^\dagger - c^\dagger a_\omega)
			+
			\int_\mathbb{R} \dd{\omega} i\hbar \sqrt{\frac{\kappa_m}{2\pi}}
			\qty(cb_\omega^\dagger - c^\dagger b_\omega).
		\end{align}
    Here, we have assumed frequency-independent (Markov approximation) and real-valued coupling rates $\kappa_c/2\pi, \kappa_m/2\pi$ to the baths.
    
   	Additionally, the cavity system is coupled to the environment through three ports, used to inject and sense microwave fields. We label them as Port 1, Port 2, and Port 3. To model the coupling of the magnon and cavity modes to the ports, we again couple them to a bosonic bath, but this time we allow the coupling rates to be complex-valued to model potential dephasing of the microwaves through the coaxial cables or due to the antenna geometry \cite{2024Bourcin}. Therefore, we assume that the cavity mode couples to Ports 1 and 2 with coupling constants $\kappa_1/2\pi,\kappa_2/2\pi$, and phases $\phi_1, \phi_2$, while the magnon couples to Port 3 only, with coupling constant $\kappa_3/2\pi$ and phase $\phi_3$. For the sake of completeness, we further consider that the cavity mode can couple to Port 3 with coupling constant $\zeta \kappa_3/2\pi$ and phase $\phi_3'$, where $\zeta$ a dimensionless constant.
    The continuous-frequency bosonic bath of port $i$ is described by creation and annihilation operators $p_{i,\omega}^\dagger, p_{i,\omega}$ with Hamiltonian
    \begin{equation}
    	H_\text{bath,ports} =
    	\sum_{i=1}^3 
    	\int_\mathbb{R} \dd{\omega} \hbar \omega p_{i,\omega}^\dagger p_{i,\omega}.
    \end{equation}
    while the interaction between the bath and the cavity system is modelled by
    \begin{equation}
    	\begin{aligned}
    		H_\text{cav--bath,ports}
    		&=
    		\int_\mathbb{R} \dd{\omega} i\hbar \sqrt{\frac{\kappa_1}{2\pi}}
    		\qty(cp_{1,\omega}^\dagger e^{-i \phi_1} - c^\dagger p_{1,\omega} e^{i \phi_1})
    		+
    		\int_\mathbb{R} \dd{\omega} i\hbar \sqrt{\frac{\kappa_2}{2\pi}}
    		\qty(cp_{2,\omega}^\dagger e^{-i \phi_2} - c^\dagger p_{2,\omega} e^{i \phi_2})\\
    		&\quad+
    		\int_\mathbb{R} \dd{\omega} i\hbar \sqrt{\frac{\kappa_3}{2\pi}}
    		\qty(mp_{3,\omega}^\dagger e^{-i \phi_3} - m^\dagger p_{3,\omega} e^{i \phi_3})
    		+
    		\int_\mathbb{R} \dd{\omega} i\hbar \sqrt{\frac{\zeta \kappa_3}{2\pi}}
    		\qty(cp_{3,\omega}^\dagger e^{-i \phi_3} - c^\dagger p_{3,\omega} e^{i \phi_3'}).
    	\end{aligned}
    \end{equation}
    
    Thus, the total Hamiltonian of the joint cavity system and environment is 
    \begin{equation}
		\begin{aligned}
			H &= H_\text{sys}
			+ H_\text{bath,intrinsic} + H_\text{cav--bath,intrinsic}
			+ H_\text{bath,ports} + H_\text{cav--bath,ports}.
		\end{aligned}
    \end{equation}

    \paragraph{Equation of motions.}
    The Heisenberg equation of motion for $p_{1,\omega}$ is 
    \begin{equation}
    	\dot{p}_{1,\omega}(t)
    	= -i \omega p_{1,\omega}(t) 
    	+ \sqrt{\frac{\kappa_1}{2\pi}} e^{-i \phi_1}c(t).
    \end{equation}
    The formal solution for an arbitrary $t_0<t$ is
    \begin{equation}
    	p_{1,\omega}(t) =
    	e^{-i \omega(t-t_0)} p_{1,\omega}(t_0)
    	+ \int_{t_0}^t \dd{t'} \sqrt{\frac{\kappa_1}{2\pi}} e^{-i \phi_1} c(t')
    	e^{-i \omega \qty(t-t')},
    \end{equation}
    and by analogy we deduce that
    \begin{equation}
    	p_{2,\omega}(t) =
    	e^{-i \omega(t-t_0)} p_{2,\omega}(t_0)
    	+ \int_{t_0}^t \dd{t'} 
    	\sqrt{\frac{\kappa_2}{2\pi}} e^{-i \phi_2} c(t') e^{-i \omega \qty(t-t')},
    \end{equation}
    \begin{equation}
		\begin{aligned}
			p_{3,\omega}(t) &=
			e^{-i \omega(t-t_0)} p_{3,\omega}(t_0)
			+ \int_{t_0}^t \dd{t'} 
			\sqrt{\frac{\kappa_3}{2\pi}} e^{-i \phi_3} m(t') e^{-i \omega \qty(t-t')}
			+ \int_{t_0}^t \dd{t'} 
			\sqrt{\frac{\zeta \kappa_3}{2\pi}} e^{-i \phi_3'} c(t') e^{-i \omega \qty(t-t')},
		\end{aligned}
    \end{equation}
	\begin{align}
    	a_{\omega}(t) &=
    	e^{-i \omega(t-t_0)} a_{\omega}(t_0)
    	+ \int_{t_0}^t \dd{t'} 
    	\sqrt{\frac{\kappa_c}{2\pi}} c(t') e^{-i \omega \qty(t-t')},\\
    	b_{\omega}(t) &=
    	e^{-i \omega(t-t_0)} b_{\omega}(t_0)
    	+ \int_{t_0}^t \dd{t'} 
    	\sqrt{\frac{\kappa_m}{2\pi}} m(t') e^{-i \omega \qty(t-t')}.
    \end{align}
    
    Taking $t_0 \to - \infty$ and using $\int_{-\infty}^t \dd{t'} f(t')\delta(t-t') = \frac{1}{2}f(t)$, the Heisenberg equation of motion for $m$ is
    \begin{align}
    	\dot{m}(t)
    	&=
    	-i \omega_m m(t) -igc(t) 
    	-\sqrt{\frac{\kappa_m}{2\pi}}
    	\int_\mathbb{R} \dd{\omega} b_{\omega}(t)
    	-\sqrt{\frac{\kappa_3}{2\pi}} e^{i \phi_3}
    	\int_\mathbb{R} \dd{\omega} p_{3,\omega}(t)\\
    	&=
    	-i \omega_m m(t) -igc(t) 
    	- \sqrt{\kappa_m} m^\text{in}(t)
    	- \frac{\kappa_m}{2\pi} \int_{t_0 \to -\infty}^t \dd{t'} m(t')
    	\int_\mathbb{R} \dd{\omega} e^{-i \omega \qty(t-t')} \nonumber\\
    	&\quad
    	- \sqrt{\kappa_3} e^{i \phi_3} p_3^\text{in}(t)
    	- \frac{\kappa_3}{2\pi} \int_{t_0 \to -\infty}^t \dd{t'} \qty(m(t') + e^{i (\phi_3-\phi_3')} \sqrt{\zeta}c(t'))
    	2\pi \delta(t-t')\\
    	&=
    	-i \widetilde{\omega}_m m(t) 
    	-\qty(ig + e^{i (\phi_3-\phi_3')} \sqrt{\zeta} \frac{\kappa_3}{2})c(t)
    	- \sqrt{\kappa_m} m^\text{in}(t)
    	- \sqrt{\kappa_3} e^{i \phi_3}  p_3^\text{in}(t),
    \end{align}
    with $\widetilde{\omega}_m = \omega_m-i\frac{\kappa_m + \kappa_3}{2}$,
    \begin{equation}
    	p_i^\text{in}(t) = \lim_{t_0 \to -\infty} \frac{1}{\sqrt{2\pi}} \int_\mathbb{R} \dd{\omega} e^{-i \omega(t-t_0)} p_{i,\omega}(t_0),
    \end{equation}
    and 
     \begin{equation}
    	c^\text{in}(t) = \lim_{t_0 \to -\infty} \frac{1}{\sqrt{2\pi}} \int_\mathbb{R} \dd{\omega} e^{-i \omega(t-t_0)} a_{\omega}(t_0),
    	\quad
    	m^\text{in}(t) = \lim_{t_0 \to -\infty} \frac{1}{\sqrt{2\pi}} \int_\mathbb{R} \dd{\omega} e^{-i \omega(t-t_0)} b_{\omega}(t_0).
    \end{equation}
    Note that we defined the input fields following the original convention of \cite{1985GardinerCollett}, but a different choice, with a minus sign instead, can also be made. This only changes the phase reference for the ports, and does not impact our results. Indeed, one can formally check that changing $p_i^{in} \mapsto -p_i^{in}$ implies $\widetilde{c} \mapsto - \widetilde{c}$ (see e.g. equation (B7)). As a consequence, one finds that $r_1 \mapsto -r_1$ and $t_2 \mapsto -t_2$, leading to a phase shift of $\pi$ which is unobservable when considering the reflection and transmission amplitudes.

    Similarly, the Heisenberg equation of motion for $c$ reads
    \begin{align}
    	\dot{c}(t)
    	&=
    	-i \omega_c c(t) -ig^* m(t) 
    	-\sqrt{\frac{\kappa_c}{2\pi}} \int_\mathbb{R} \dd{\omega} a_{\omega}(t) \nonumber\\
    	&\quad
    	-\sqrt{\frac{\kappa_1}{2\pi}} e^{i \phi_1} \int_\mathbb{R} \dd{\omega} p_{1,\omega}(t)
    	-\sqrt{\frac{\kappa_2}{2\pi}} e^{i \phi_2} \int_\mathbb{R} \dd{\omega} p_{2,\omega}(t)
    	-\sqrt{\frac{\zeta \kappa_3}{2\pi}} e^{i \phi_3'} \int_\mathbb{R} \dd{\omega} p_{3,\omega}(t)\\
    	&=
    	-i \omega_c c(t) -ig^* m(t) - e^{i (\phi_3'-\phi_3)} \sqrt{\zeta} \frac{\kappa_3}{2}m(t)
    	-\frac{\kappa_c+\kappa_1+\kappa_2 + \zeta \kappa_3}{2} c(t) \nonumber\\
    	&\quad
    	-\sqrt{\kappa_c} c^\text{in}(t)
    	-\sqrt{\kappa_1} e^{i \phi_1} p_1^\text{in}(t)
    	-\sqrt{\kappa_2} e^{i \phi_2} p_2^\text{in}(t)
    	-\sqrt{\zeta \kappa_3} e^{ \phi_3'} p_3^\text{in}(t)\\
    	&=
    	-i \widetilde{\omega}_c c(t) 
    	-\widetilde{g}' m(t)
    	-\sqrt{\kappa_c} c^\text{in}(t)
    	-\sqrt{\kappa_1} e^{i \phi_1} p_1^\text{in}(t)
    	-\sqrt{\kappa_2} e^{i \phi_2} p_2^\text{in}(t)
    	-\sqrt{\zeta \kappa_3} e^{i \phi_3'} p_3^\text{in}(t).
    \end{align}
    where we defined $\widetilde{\omega}_c = \omega_c-i\frac{\kappa_c + \kappa_1+\kappa_2 + \zeta \kappa_3}{2}$ and $\widetilde{g}' = g^* -i \sqrt{\zeta} \frac{\kappa_3}{2} e^{i (\phi_3'-\phi_3)}$.
    Similarly defining $\widetilde{g} = g -i \sqrt{\zeta} \frac{\kappa_3}{2} e^{i (\phi_3-\phi_3')}$, we can rewrite the Heisenberg equation of motions as
    \begin{align}
    	\dot{m} &=
    	-i\widetilde{\omega}_m m -\widetilde{g}c
    	-\sqrt{\kappa_m} m^\text{in}(t)
    	-\sqrt{\kappa_3} e^{i \phi_3} p_3^\text{in}(t),\\
    	\dot{c} &= 
    	-i\widetilde{\omega}_c c - i\widetilde{g}' m
    	-\sqrt{\kappa_c} c^\text{in}(t)
    	-\sqrt{\kappa_1} e^{i \phi_1} p_1^\text{in}(t)
    	-\sqrt{\kappa_2} e^{i \phi_2} p_2^\text{in}(t)
    	-\sqrt{\zeta \kappa_3} e^{i \phi_3'} p_3^\text{in}(t),
    \end{align}
    which are the QLEs given in the main text.

    \section{\label{app:input-output}Calculation of the reflection at Port 1}
    \paragraph{Frequency-space solutions.}
    We now perform a semi-classical approximation, and replace operators by their expectation value. 
    Recalling that $\expval{c^\text{in}} = \expval{m^\text{in}} = 0$, the QLEs simplify to
    \begin{align}
    	\dot{m} &=
    	-i\widetilde{\omega}_m m -i \widetilde g c
    	-\sqrt{\kappa_3} e^{i \phi_3} p_3^\text{in}(t),\\
    	\dot{c} &= 
    	-i\widetilde{\omega}_c c -i \widetilde g' m
    	-\sqrt{\kappa_1} e^{i \phi_1} p_1^\text{in}(t)
    	-\sqrt{\kappa_2} e^{i \phi_2} p_2^\text{in}(t)
    	-\sqrt{\zeta \kappa_3} e^{i \phi_3'} p_3^\text{in}(t),
    \end{align}
    and in frequency space,
    \begin{gather}
    	-i \omega \widetilde{m} =
    	-i\widetilde{\omega}_m \widetilde{m} -i\widetilde g \widetilde{c}
    	-\sqrt{\kappa_3} e^{i \phi_3} \widetilde p_3^\text{in},\\
    	-i\widetilde{\Delta}_c \widetilde{c} 
    	=
    	-i \widetilde g' \widetilde{m}
    	-\sqrt{\kappa_1} e^{i \phi_1} \widetilde p_1^\text{in}
    	-\sqrt{\kappa_2} e^{i \phi_2} \widetilde p_2^\text{in}
    	-\sqrt{\zeta \kappa_3} e^{i \phi_3'} \widetilde p_3^\text{in},
    \end{gather}
    with $\widetilde{\Delta}_c=\omega-\widetilde{\omega}_c$ and $\widetilde{\Delta}_m=\omega- \widetilde{\omega}_m$.
    In particular, the first equation gives
    \begin{gather}
    	\label{eq:app:temp-m}
    	\widetilde{m}
    	=
    	\frac{\widetilde g \widetilde{c} - i\sqrt{\kappa_3} e^{i \phi_3}\widetilde{p}_3^\text{in}}{\widetilde{\Delta}_m},
    \end{gather}
    which inserted in the second equation leads to
    \begin{gather}
    	\widetilde{c} 
    	=
    	\frac{
    		\widetilde g' \widetilde{m}
    		-i\sqrt{\kappa_1} e^{i \phi_1} \widetilde{p}_1^\text{in}
    		-i\sqrt{\kappa_2} e^{i \phi_2} \widetilde{p}_2^\text{in}
    		-i\sqrt{\zeta \kappa_3} e^{i \phi_3'} \widetilde{p}_3^\text{in}
    	}{\widetilde{\Delta}_c}\\
    	\label{eq:app:temp-c}
    	\widetilde{c} 
    	=
    	\frac{-i}{A(\omega)} \qty[
	    	\widetilde{\Delta}_m \sqrt{\kappa_1} e^{i \phi_1} \widetilde{p}_1^\text{in}
    		+\widetilde{\Delta}_m \sqrt{\kappa_2} e^{i \phi_2} \widetilde{p}_2^\text{in}
    		+\qty(\widetilde g' e^{i \phi_3} + \sqrt{\zeta} e^{i \phi_3'}\widetilde{\Delta}_m)\sqrt{\kappa_3} \widetilde{p}_3^\text{in}
    	]
    \end{gather}
    where $A(\omega) = \widetilde{\Delta}_c \widetilde{\Delta}_m - \widetilde g \widetilde g'$.

    \paragraph{Input-output relations.}
    To derive the input-output relations, we first write the formal solution of $p_{1,\omega}$ for $t< t_1$ as
    \begin{equation}
    	p_{1,\omega}(t) =
    	e^{-i \omega(t-t_1)} p_{1,\omega}(t_1)
    	- \int_{t}^{t_1} \dd{t'} \sqrt{\frac{\kappa_1}{2\pi}} e^{-i \phi_1} c(t') 
    	e^{-i \omega \qty(t-t')}.
    \end{equation}
    and we have on the one hand
    \begin{equation}
    	\int_\mathbb{R} \dd{\omega} p_{1,\omega}(t)
    	= \sqrt{2\pi} \qty(
    		p_1^\text{in}(t)
    		+ \frac{\sqrt{\kappa_1}}{2} e^{-i \phi_1} c(t)
    	) ,
    \end{equation}
    and on the other
    \begin{equation}
    	\int_\mathbb{R} \dd{\omega} p_{1,\omega}(t)
    	= \sqrt{2\pi} \qty(
    		p_{1,\omega}^\text{out}(t)
    		- \frac{\sqrt{\kappa_1}}{2} e^{-i \phi_1} c(t) 
    	),
    \end{equation}
    so that the input-output relation is
    \begin{equation}
    	\label{eq:input-output-relation}
    	p_1^\text{out}(t) = p_1^\text{in}(t) + \sqrt{\kappa_1} e^{-i \phi_1} c(t).
    \end{equation}
    We can similarly derive
    \begin{align}
    	p_2^\text{out}(t) &= p_2^\text{in}(t) + \sqrt{\kappa_2} e^{-i \phi_2} c(t),\\
    	p_3^\text{out}(t) &= p_3^\text{in}(t) + \sqrt{\kappa_3} e^{-i \phi_3} m(t) 
    	+ \sqrt{\zeta \kappa_3} e^{-i \phi_3'} c(t).
    \end{align}

    \paragraph{Modelling coherent drives.}
    To examine the reflection and transmission, we consider that $p_1$, coupling to the cavity mode, is a coherent drive at frequency $\omega_d/2\pi$. Such a coherent drive is expected to reproduce the classical dynamics through the use of coherent states \cite{2006Fox}. Formally, the state space for $p_1$ is the product of the Fock space of the bath operators $p_{1,\omega}$ for each frequency $\omega/2\pi$. Formally, the state of the bath is the tensor product $\bigotimes_{\omega \in \mathbb R} \ket{\mathcal \psi_{\omega}}$, where $\ket{\mathcal \psi_{\omega}}$ is the state of the bosonic mode described by the annihilation operator $p_{1,\omega}$. If we assume a coherent drive, then only the mode of frequency $\omega_d/2\pi$ has a non-vanishing number of excitation, which we take to be a coherent state $\ket{\mathcal E_{\omega_d}}$. Formally, $\ket{\psi_{\omega}} = \ket{0}$ if $\omega \neq\omega_d$, and $\ket{\psi_{\omega}} = \ket{\mathcal E_{\omega_d}}$ if $\omega = \omega_d$. Hence, $\expval{p_1} = \mel{\mathcal E_{\omega_d}}{p_1}{\mathcal E_{\omega_d}} = \mathcal E_{\omega_d} \in \mathbb{C}$, since we recall that coherent states are eigenstates of annihilation operators \cite{2006Fox}. We make a similar approximation for Port 3 driving the magnon, albeit with a different amplitude and phase, and hence $\expval{p_3} = \delta e^{i \phi} \mathcal E_{\omega_d}$.
    
    A convenient choice for the coherent state $\mathcal E_{\omega_d}$ is $\mathcal E_{\omega_d}=\lim_{t_0 \to -\infty} \alpha e^{-i \omega_d t_0}$ with $\mathcal E \in \mathbb R$. Indeed, in the time-domain this gives for instance
    \begin{align}
    	\expval{p_1^\text{in}(t)}
    	&= 
    	\mel{\mathcal E_{\omega_d}}{p_1^\text{in}(t)}{\mathcal E_{\omega_d}} \\
    	&= 
    	\lim_{t_0 \to -\infty} \mel{
    		\alpha e^{i \omega_d t_0}
    		}{\frac{1}{\sqrt{2\pi}} \int_\mathbb{R} \dd{\omega} e^{-i \omega(t-t_0)} p_{1,\omega}(t_0)
    		}{\mathcal E e^{-i \omega_d t_0}
    	}\\
    	&= 
    	\lim_{t_0 \to -\infty} \mel{
    		\mathcal E e^{i \omega_d t_0}
	    	}{\frac{1}{\sqrt{2\pi}} e^{-i \omega_d(t-t_0)} \alpha e^{-i \omega_d t_0}
    		}{\mathcal E e^{-i \omega_d t_0}
    	}\\
    	&= \frac{\mathcal E}{\sqrt{2\pi}} e^{-i \omega_d t}
    \end{align}
    where in the third line we used the fact that a coherent state is an eigenstate of the annihilation operator, and that $p_{1,\omega} \ket{\mathcal E e^{-i \omega_d t_0}} = 0$ if $\omega \neq \omega_d$.

    \paragraph{Expression of the reflection coefficient.}
    From $\expval{p_1^\text{in}(t)} =\frac{\mathcal E}{\sqrt{2\pi}} e^{-i \omega_d t}$, we see that $\expval{\widetilde p_1^\text{in}(\omega)} = \mathcal E \delta(\omega-\omega_d)$. Thus, assimilating $\omega$ to the frequency of the drive, the reflection at Port 1 reads
    \begin{align}
    	r_{1}(\omega) 
    	&=
    	\left. \frac{\widetilde p_1^\text{out}}{\widetilde p_1^\text{in}}\right|_{\widetilde p_2^\text{in}=0}
    	= 1+
    	\sqrt{\kappa_1} e^{-i \phi_1} 
    	\frac{\left.\widetilde c\right|_{\widetilde p_2^\text{in}=0}}{\widetilde p_1^\text{in}}\\
    	&=
    	1 
    	-i \sqrt{\kappa_1}\frac{\widetilde{\Delta}_m \sqrt{\kappa_1}}{A(\omega)} 
	    -i \sqrt{\kappa_1} e^{-i \phi_1} 
	    \frac{\qty(\widetilde g' e^{i \phi_3} + \sqrt{\zeta} e^{i \phi_3'}\widetilde{\Delta}_m)\sqrt{\kappa_3}}{A(\omega)} 
	    \frac{\widetilde p_3^\text{in}}{\widetilde p_1^\text{in}}\\
	    \label{eq:reflection:crosstalk}
    	&=
    	1 
    	-i \sqrt{\kappa_1}\frac{
    		\widetilde{\Delta}_m \sqrt{\kappa_1}
    		+ \qty(\widetilde g' e^{i(\phi_3- \phi_1)} + \sqrt{\zeta} e^{i (\phi_3'-\phi_1)}\widetilde{\Delta}_m)\sqrt{\kappa_3} \delta e^{i \phi}
    	}{A(\omega)}
    \end{align}
    Notice that for $\kappa_3=0$, i.e. only the cavity mode is driven, the expression for the reflection reduces to the $S_{11}$ parameter.

    \paragraph{Normalisation.} 
    For a standard coherent state $\ket{\mathcal E}$, the mean number of particle is given by $\abs{\mathcal E}^2$. Here, the unit of the ports $\expval{p_k}$ are $\sqrt{\omega}$, which can be seen by considering the quantum Langevin equations. Hence, for the ports, $\abs{\mathcal E}^2$ is a number of photons per second, which can be linked with the power of the drive $P$ by
    \begin{equation}
    	\mathcal E = \sqrt{\frac{P}{\hbar \omega_d}}.
    \end{equation}
    Given that $\expval{p_3} = \delta e^{i\phi} \expval{p_1}$, we deduce that the power difference between Port 1 and Port 3 is given by $\delta^2$, which allows to renormalise the expression of the reflection and transmission.

    \section{\label{app:R1}Expression of the reflection at Port 1 with standard S-parameters}
    In this note, we derive the expressions of the standard S-parameters, when only one port is active at a time. Assuming that port 3 is not active, i.e. $\expval{p_3^\text{in}} =0$ (which corresponds to $\delta=0$), the reflection $r_{1}$ and transmission $t_{2}$ coefficients of \cref{eq:reflection:crosstalk,eq:transmission:crosstalk} reduce to 
    \begin{align}
    	S_{11}(\omega)
    	&=
    	1 
    	-i \frac{
    		\widetilde{\Delta}_m \kappa_1
    	}{\widetilde{\Delta}_c\widetilde{\Delta}_m - \widetilde g \widetilde g'},\\
    	\label{eq:S21}
    	S_{21}(\omega)
    	&= 
    	-i
    	\frac{
    		\widetilde{\Delta}_m\sqrt{\kappa_1 \kappa_2} e^{i(\phi_1- \phi_2)}
    	}{\widetilde{\Delta}_c\widetilde{\Delta}_m - \widetilde g \widetilde g'}.
    \end{align}
    which are the standard S parameters for a two-port cavity. Note that by symmetry we also have
    \begin{equation}
    	S_{12}(\omega)
    	= 
    	-i
    	\frac{
    		\widetilde{\Delta}_m\sqrt{\kappa_1 \kappa_2} e^{i(\phi_2- \phi_1)}
    	}{\widetilde{\Delta}_c\widetilde{\Delta}_m - \widetilde g \widetilde g'},
    	\quad
    	S_{22}(\omega)
    	=
    	1 
    	-i \frac{
    		\widetilde{\Delta}_m \kappa_2
    	}{\widetilde{\Delta}_c\widetilde{\Delta}_m - \widetilde g \widetilde g'}.
    \end{equation}
    
    Similarly, using \cref{eq:app:temp-c} we find
    \begin{align}
		\label{eq:S13}
    	S_{13}(\omega) 
    	&=
    	\left. \frac{\widetilde p_1^\text{out}}{\widetilde p_3^\text{in}}\right|_{\widetilde p_1^\text{in}=\widetilde p_2^\text{in}=0} 
    	= 
    	\sqrt{\kappa_1} e^{-i \phi_1} 
    	\frac{
    		\left.
    		\widetilde c
    		\right|_{\widetilde p_1^\text{in}=\widetilde p_2^\text{in}=0}
    	}{\widetilde p_3^\text{in}}
    	=
    	-i \sqrt{\kappa_1 \kappa_3}\frac{
    		\widetilde g' e^{i(\phi_3- \phi_1)} + \sqrt{\zeta} e^{i (\phi_3'-\phi_1)}\widetilde{\Delta}_m
    	}{A(\omega)},
    \end{align}
	and hence
	\begin{equation}
		r_1(\omega) = S_{11} + \delta e^{i \phi}S_{13} = 1 - i \sqrt{\kappa_1}
		\frac{
			\widetilde{\Delta}_m \sqrt{\kappa_1} 
			+ 
			\qty(\widetilde g' e^{i(\phi_3- \phi_1)} + \sqrt{\zeta} e^{i (\phi_3'-\phi_1)}\widetilde{\Delta}_m)
			\sqrt{\kappa_3} \delta e^{i \phi}
		}{A(\omega)}
	\end{equation}
	which matches with the equation given in the main text for $\zeta=0$.

	\section{\label{app:transmission}Calculation of the transmission}
	Using the results of the input-output theory above, the transmission through Port 2 when Port 1 and 3 are active is calculated to be
	\begin{align}
		t_{2}(\omega) 
		&=
		\left. \frac{\widetilde p_2^\text{out}}{\widetilde p_1^\text{in}}\right|_{\widetilde p_2^\text{in}=0}
		= \sqrt{\kappa_2} e^{-i \phi_2} 
		\frac{\left.\widetilde c\right|_{\widetilde p_2^\text{in}=0}}{\widetilde p_1^\text{in}}\\
		\label{eq:transmission:crosstalk}
		&= 
		-i\sqrt{\kappa_2}
		\frac{\widetilde{\Delta}_m \sqrt{\kappa_1}  e^{i(\phi_1- \phi_2)} + \qty(\widetilde g' e^{i(\phi_3- \phi_2)} + \sqrt{\zeta} e^{i (\phi_3'-\phi_2)}\widetilde{\Delta}_m)\sqrt{\kappa_3}\delta e^{i \phi}}{A(\omega)}.
	\end{align}
	Thus, the normalised transmission at Port 2 is $T_2 = \frac{1}{\sqrt{1+\delta^2}} t_{2}$.

	\paragraph{Expression with S parameters.}
	Furthermore, from \cref{eq:S13}, we see that $S_{23}$ can be obtained from $S_{13}$ by replacing $(\kappa_1, \phi_1) \mapsto (\kappa_2, \phi_2)$, and thus we have
	    \begin{align}
	    \label{eq:S23}
		S_{23}(\omega) 
		&=
		\left. \frac{\widetilde p_2^\text{out}}{\widetilde p_3^\text{in}}\right|_{\widetilde p_1^\text{in}=\widetilde p_2^\text{in}=0} 
		= 
		\sqrt{\kappa_2} e^{-i \phi_2} 
		\frac{
			\left.
			\widetilde c
			\right|_{\widetilde p_1^\text{in}=\widetilde p_2^\text{in}=0}
		}{\widetilde p_3^\text{in}}
		=
		-i \sqrt{\kappa_2 \kappa_3}\frac{
			\widetilde g' e^{i(\phi_3- \phi_2)} + \sqrt{\zeta} e^{i (\phi_3'-\phi_2)}\widetilde{\Delta}_m
		}{A(\omega)}.
	\end{align}
	Hence, comparing \cref{eq:transmission:crosstalk} with \cref{eq:S21,eq:S23} we conclude that $t_2 = S_{21} + \delta e^{i\phi} S_{23}$.

	\paragraph{Zeros of the nominator.}
	The zeros of the nominator of $T_2$ are given by 
	\begin{gather}
		(\omega - \widetilde \omega_m) \qty[
			\sqrt{\kappa_1}  e^{i(\phi_1- \phi_2)} +
			\sqrt{\zeta \kappa_3} e^{i (\phi_3'-\phi_2)} \delta e^{i \phi}
		]
		+ \qty(g^* -i \sqrt{\zeta} \frac{\kappa_3}{2} e^{i (\phi_3'-\phi_3)}) e^{i (\phi_3-\phi_2)} \sqrt{\kappa_3} \delta e^{i \phi}
		= 0
	\end{gather}
	i.e.
	\begin{align}
		\omega 
		&= 
		\widetilde{\omega}_m
		+ 
		\frac{
			\qty(g^* -i \sqrt{\zeta} \frac{\kappa_3}{2} e^{i (\phi_3'-\phi_3)}) e^{i \phi_3} \sqrt{\kappa_3} \delta e^{i \phi}
		}{
			\sqrt{\kappa_1}  e^{i\phi_1} +
		\sqrt{\zeta \kappa_3} e^{i \phi_3'} \delta e^{i \phi}
		}.
	\end{align}

	\section{\label{app:R2}Expression of the reflection at Port 2 and the transmission at Port 1}
	In the main text, we only consider that Port 1 is being driven, while Port 2 is passive and used as a probe. In this note, we exchange the role of Port 1 and Port 2 to check that the results are identical. Thus, we now have $\expval{p_2^\text{in}(t)} =\frac{\mathcal E}{\sqrt{2\pi}} e^{-i \omega_d t}$, $\expval{p_3^\text{in}} = \delta e^{i \phi} \expval{p_1^\text{in}}$, and $\expval{p_2^\text{in}}=0$. The reflection at Port 2 is
	\begin{align}
		r_{2}(\omega) 
		&=
		\left. \frac{\widetilde p_2^\text{out}}{\widetilde p_2^\text{in}}\right|_{\widetilde p_1^\text{in}=0}
		= 1+
		\sqrt{\kappa_2} e^{-i \phi_2} 
		\frac{\left.\widetilde c\right|_{\widetilde p_1^\text{in}=0}}{\widetilde p_2^\text{in}}\\
		&=
		1 
		-i \sqrt{\kappa_2}\frac{
			\widetilde{\Delta}_m \sqrt{\kappa_2}
			+ \qty(\widetilde g' e^{i(\phi_3- \phi_2)} + \sqrt{\zeta} e^{i (\phi_3'-\phi_2)}\widetilde{\Delta}_m)\sqrt{\kappa_3} \delta e^{i \phi}
		}{A(\omega)},
	\end{align}
	and hence it is formally identical to $r_1$ after replacing the index 1 by 2. For the transmission through Port 1,
	\begin{align}
		t_{1}(\omega) 
		&=
		\left. \frac{\widetilde p_1^\text{out}}{\widetilde p_2^\text{in}}\right|_{\widetilde p_1^\text{in}=0}
		= \sqrt{\kappa_1} e^{-i \phi_1} 
		\frac{\left.\widetilde c\right|_{\widetilde p_1^\text{in}=0}}{\widetilde p_2^\text{in}}\\
		&= 
		-i\sqrt{\kappa_1}
		\frac{\widetilde{\Delta}_m \sqrt{\kappa_2}  e^{i(\phi_2- \phi_1)} + \qty(\widetilde g' e^{i(\phi_3- \phi_1)} + \sqrt{\zeta} e^{i (\phi_3'-\phi_1)}\widetilde{\Delta}_m)\sqrt{\kappa_3}\delta e^{i \phi}}{A(\omega)}.
	\end{align}

	\section{\label{app:crosstalk}Effective coupling strength and crosstalk}
	For non-vanishing coupling of Port 3 to the cavity mode, $\zeta \neq 0$, we recall that the reflection coefficient given by \cref{eq:reflection:crosstalk} is
	\begin{align}
		r_{1}(\omega) 
		&= \frac{N(\omega)}{D(\omega)}
		=
		1 
		-i \sqrt{\kappa_1}\frac{
			\widetilde{\Delta}_m \sqrt{\kappa_1}
			+ \qty(\widetilde g' e^{i(\phi_3 - \phi_1)} + \sqrt{\zeta} e^{i(\phi_3'-\phi_1)}\widetilde{\Delta}_m)\sqrt{\kappa_3} \delta e^{i \phi}
		}{A(\omega)}\\
		&=
		\frac{
			\widetilde{\Delta}_c \widetilde{\Delta}_m 
			- \widetilde g \widetilde g'
			-i\kappa_1 \widetilde{\Delta}_m
			-i\widetilde g' \sqrt{\kappa_1 \kappa_3} e^{i(\phi_3 - \phi_1)} \delta e^{i \phi}
			-i\widetilde \Delta_m \sqrt{\zeta} \sqrt{\kappa_1 \kappa_3} e^{i(\phi_3'-\phi_1)} \delta e^{i \phi}
		}{
			\widetilde{\Delta}_c \widetilde{\Delta}_m 
			- \widetilde g \widetilde g'
		}
		.
	\end{align}
	The nominator is
	\begin{align}
		N(\omega)
		&=
		\omega^2 
		- \qty(\widetilde \omega_c+ \widetilde \omega_m) \omega
		+ \widetilde \omega_c \widetilde \omega_m - \widetilde g \widetilde g'
		-i\kappa_1 \qty(\omega-\widetilde \omega_m)
		-i \widetilde g'\sqrt{\kappa_1 \kappa_3} \delta e^{i (\phi+\phi_3-\phi_1)}
		-i \qty(\omega-\widetilde \omega_m) \sqrt \zeta \sqrt{\kappa_1 \kappa_3} \delta e^{i (\phi+\phi_3'-\phi_1)}\\
		&=
		\omega^2
		- \qty(
		\widetilde \omega_c 
		+ i \qty[\kappa_1 + \sqrt \zeta \sqrt{\kappa_1 \kappa_3} \delta e^{i (\phi+\phi_3'-\phi_1)}] 
		+ \widetilde \omega_m
		) \omega
		+ 
		\qty(
		\widetilde \omega_c
		+i \qty[\kappa_1+\sqrt \zeta \sqrt{\kappa_1 \kappa_3} \delta e^{i (\phi+\phi_3'-\phi_1)}]
		)\widetilde \omega_m \\
		&\quad 
		-\widetilde g \widetilde g'
		-i\widetilde g'\sqrt{\kappa_1 \kappa_3} \delta e^{i (\phi+\phi_3-\phi_1)} \nonumber\\
		&=
		\omega^2
		- \qty(\widetilde \omega_c'' + \widetilde \omega_m) 
		\omega
		+ 
		\widetilde \omega_c'' \widetilde \omega_m
		- \widetilde{G}^2
	\end{align}
	where $\omega_c'' = \omega_c + i \qty[\kappa_1+\sqrt \zeta \sqrt{\kappa_1 \kappa_3} \delta e^{i (\phi+\phi_3'-\phi_1)}]$ and, recalling that $g = \abs{g}e^{i \varphi}$,
	\begin{align}
		\widetilde G 
		&=
		\sqrt{\widetilde g \widetilde g' + i\widetilde g'\sqrt{\kappa_1 \kappa_3} \delta e^{i (\phi+\phi_3-\phi_1)}}\\
		&=
		\sqrt{
			\qty(g-i\sqrt \zeta e^{i(\phi_3-\phi_3')} \frac{\kappa_3}{2})
			\qty(g^*-i\sqrt \zeta e^{-i(\phi_3-\phi_3')} \frac{\kappa_3}{2}) 
			+ i \qty(g^*-i\sqrt \zeta e^{-i(\phi_3-\phi_3')} \frac{\kappa_3}{2}) \sqrt{\kappa_1 \kappa_3} \delta e^{i (\phi+\phi_3-\phi_1)}
		}\\
		\label{eq:coupling-strength-crosstalk}
		&=
		\abs{g} \sqrt{
			1 
			+ \frac{\sqrt{\kappa_1 \kappa_3}}{\abs{g}} \delta e^{i (\phi+\phi_3-\phi_1 + \frac{\pi}{2}-\varphi)}
			- \frac{\sqrt \zeta \kappa_3}{\abs{g}} \qty(
				\frac{\sqrt \zeta}{4} \frac{\kappa_3}{\abs{g}}
				+i\cos(\varphi + \phi_3'-\phi_3)
				- \frac{\sqrt{\kappa_1 \kappa_3}}{2 \abs{g}}  \delta e^{i (\phi+\phi_3'-\phi_1)}
			)
		}.
	\end{align}
	
\begin{figure}[t]
		\centering
		\includegraphics[width=0.49\linewidth]{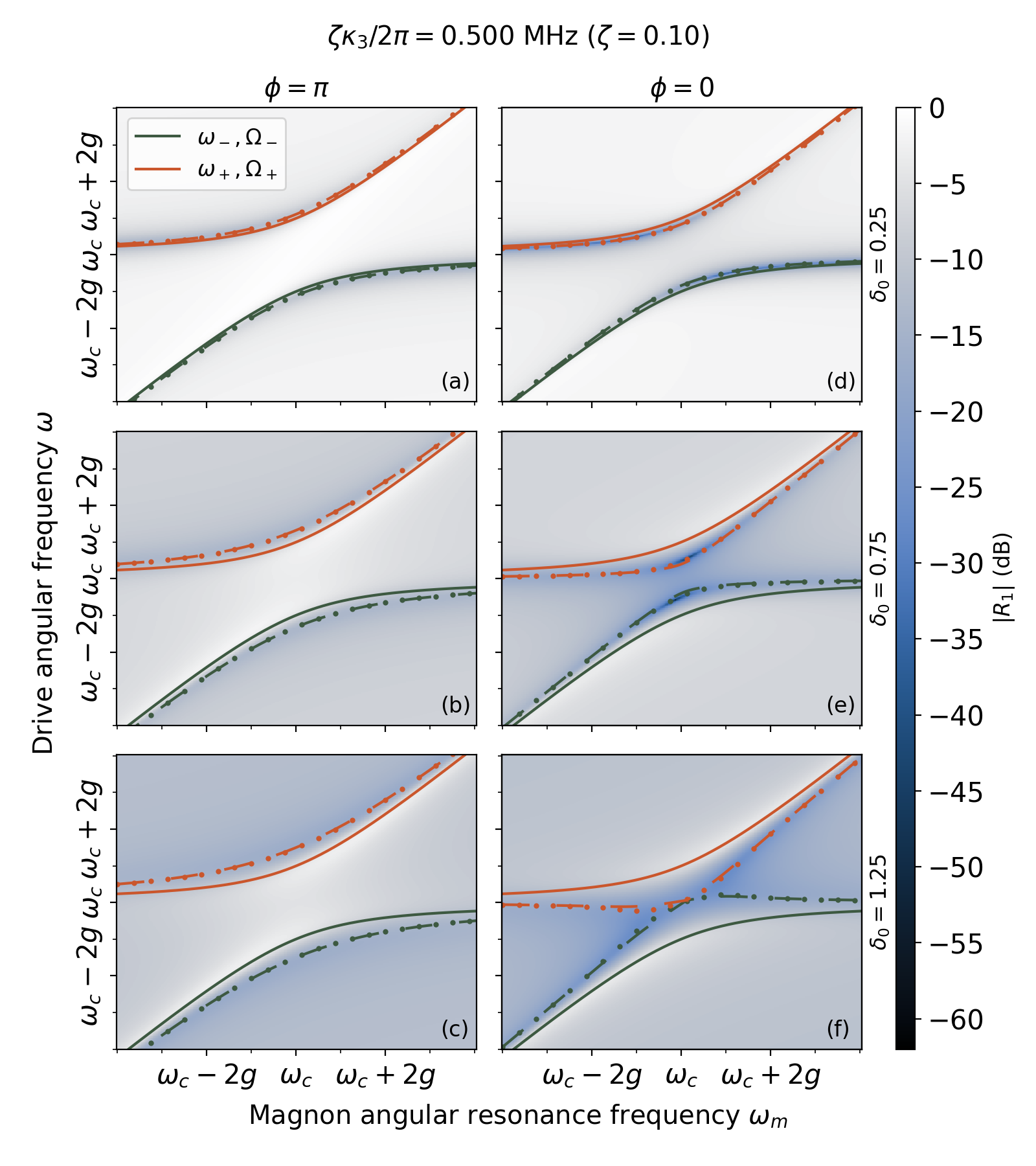}
		\includegraphics[width=0.49\linewidth]{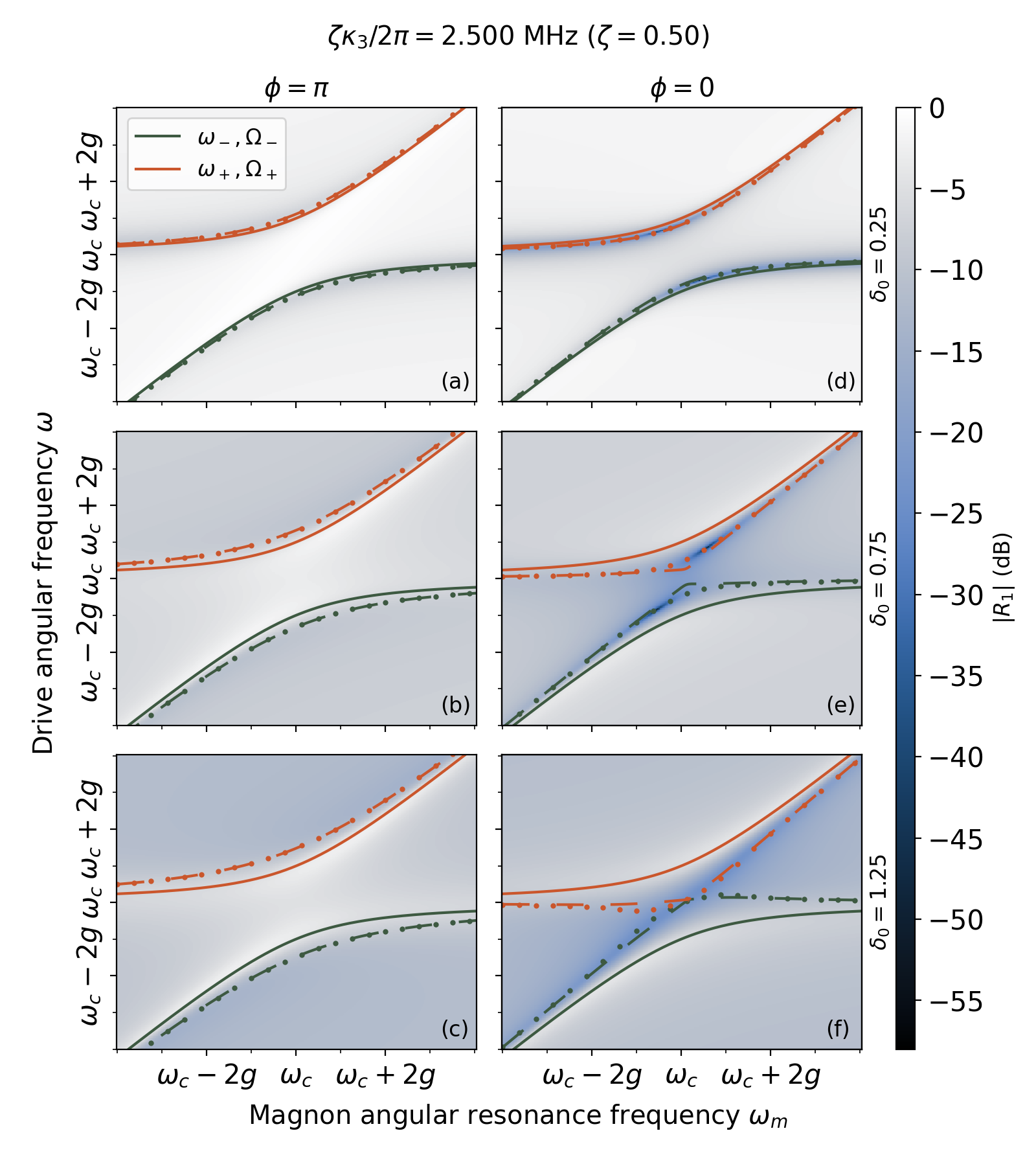}
		\vspace{-2em}
		\phantomsubfloat{\label{fig:reflection-zeta-0.1}}
		\phantomsubfloat{\label{fig:reflection-zeta-0.5}}
		\caption{Plot of the reflection amplitude $\abs{R_1}$ for $\zeta=0.1$ (left) and $\zeta=0.5$ (right), with other parameters as in figure 2 of the main text. The dashed lines corresponds to $\zeta\neq0$ (they are exactly the same as in figure 2 of the main text) while the dots corresponds to $\zeta = 0$. }
		\label{fig:reflection-zeta}
	\end{figure}

	The first two terms under the square root correspond to the formula given in the main text, which corresponds to $\zeta=0$. 
	In \cref{fig:reflection-zeta} we plot the reflection coefficient with parameters identical to those used in the main text, where $\kappa_3/g = 1/2$.

    \section{\label{app:two-tone-coherent-drive}Simplification of the effective Hamiltonian}
    The effective Hamiltonian in the main text reads
    \begin{equation}
   		H_\text{eff}
   		=
   		\hbar \widetilde \omega_c c^\dagger c 
   		+ \hbar \widetilde \omega_m m^\dagger m
   		+ \hbar g \qty(cm^\dagger + c^\dagger m)
   		+i\hbar\qty(\mathcal E_1 c e^{i\omega t} - \hc)
   		+i\hbar \qty(\mathcal E_3 m e^{i\omega t}- \hc).
    \end{equation}
    To simplify the calculations, we define $\epsilon_c = i \mathcal E_1$ and  $\epsilon_m = i \mathcal E_3$, and we obtain
    \begin{equation}
    	H_\text{eff}
    	=
    	\hbar \widetilde \omega_c c^\dagger c 
    	+ \hbar \widetilde \omega_m m^\dagger m
    	+ \hbar g \qty(cm^\dagger + c^\dagger m)
    	+\hbar \qty(\epsilon_c c e^{i\omega t} + \epsilon_c^* c^\dagger e^{-i\omega t})
    	+\hbar \qty(\epsilon_m m e^{i\omega t} + \epsilon_m^* m^\dagger e^{-i\omega t}).
    \end{equation}
    In a frame rotating with the drive, which corresponds to the unitary transformation $U=\exp\qty[i \omega_d t \qty(c^\dagger c + m^\dagger m)]$, we obtain
    \begin{equation}
    	H_\text{eff} = \hbar \widetilde \Delta_c c^\dagger c 
    	+ \hbar \widetilde \Delta_m m^\dagger m 
    	+ \hbar g\qty(cm^\dagger + c^\dagger m)
    	+ \hbar \qty(\epsilon_c^* c + \epsilon_c^* c^\dagger) 
    	+ \hbar \qty(\epsilon_m^* m + \epsilon_m^* m^\dagger),
    	\label{eq:two-tone-driving:hamiltonian}
    \end{equation}
    with the detunings $\widetilde \Delta_c = \widetilde \omega_c - \omega_d$, $\widetilde \Delta_m = \widetilde \omega_m - \omega_d$.

    Let us consider the driving terms as a perturbation $V$, and split the Hamiltonian of \cref{eq:two-tone-driving:hamiltonian} into $H = H_0+V$ where
    \begin{align}
    	H_0 &= \hbar \Delta_c c^\dagger c + \Delta_m m^\dagger m + \hbar g\qty(cm^\dagger + c^\dagger m),\\
    	V
    	&= \hbar\qty(\epsilon_c c + \epsilon_c^* c^\dagger) + \hbar\qty(\epsilon_m m + \epsilon_m^* m^\dagger).
    \end{align}
    We try to find a Schrieffer-Wolf transformation \cite{1966Schrieffer} $U=e^\Lambda$ with $\Lambda = (\lambda_c c- \lambda_c^* c^\dagger) + (\lambda_m m- \lambda_m^* m^\dagger)$, and $\lambda_c,\lambda_m$ constants to determine such that $V + \qty[\Lambda, H_0] =0$. Note that $\Lambda^\dagger = - \Lambda$, such that $e^\Lambda$ is unitary. We have
    \begin{align}
    	V + \qty[\Lambda, H_0] 
    	&= 
    	\epsilon_c c + \epsilon_c^* c^\dagger + \epsilon_m m + \epsilon_m^* m^\dagger \nonumber\\
    	&\quad
    	+ \qty[\lambda_c c - \lambda_c^*c^\dagger, \hbar \Delta_c c^\dagger c]
    	+ \qty[\lambda_c c - \lambda_c^*c^\dagger, \hbar g\qty(cm^\dagger + c^\dagger m)] \nonumber\\
    	&\quad
    	+ \qty[\lambda_m m - \lambda_m^*m^\dagger, \hbar \Delta_m m^\dagger m]
    	+ \qty[\lambda_m m - \lambda_m^*m^\dagger, \hbar g\qty(cm^\dagger + c^\dagger m)]\\
    	&=
    	\epsilon_c c + \epsilon_c^* c^\dagger + \epsilon_m m + \epsilon_m^* m^\dagger \nonumber\\
    	&\quad
    	+ \hbar \Delta_c \qty(\lambda_c c + \lambda_c^* c^\dagger)
    	+ \hbar g \qty(\lambda_c m + \lambda_c^*m^\dagger) \nonumber\\
    	&\quad
    	+ \hbar \Delta_m \qty(\lambda_m m + \lambda_m^*m^\dagger)
    	+ \hbar g \qty(\lambda_m c + \lambda_m^*c^\dagger)\\
    	&=
    	\hbar \qty[
    	\qty(\epsilon_c + \Delta_c \lambda_c + g \lambda_m)c
    	+ \qty(\epsilon_c^* + \Delta_c \lambda_c^* + g \lambda_m^*)c^\dagger
    	] \\
    	&\quad
    	+ \hbar \qty[
    	\qty(\epsilon_m + g \lambda_c + \Delta_m \lambda_m)m
    	+ \qty(\epsilon_m^* + g \lambda_c^* + \Delta_m \lambda_m^*)m^\dagger
    	]\nonumber.
    \end{align}
    Imposing $V + \qty[\Lambda, H_0] = 0$ (Schrieffer-Wolff condition) leads to 
    \begin{equation}
    	\begin{cases}
    		\epsilon_c + \Delta_c \lambda_c + g \lambda_m = 0\\
    		\epsilon_c^* + \Delta_c \lambda_c^* + g \lambda_m^* = 0\\
    		\epsilon_m + g \lambda_c + \Delta_m \lambda_m = 0\\
    		\epsilon_m^* + g \lambda_c^* + \Delta_m \lambda_m^* = 0
    	\end{cases},
    \end{equation}
    where we recognise that we get twice the same constraints by hermiticity. We can reduce the system to
    \begin{equation}
    	\begin{cases}
    		\epsilon_c + \Delta_c \lambda_c + g \lambda_m = 0\\
    		\epsilon_m + g \lambda_c + \Delta_m \lambda_m = 0\\
    	\end{cases}, \quad
    	\begin{cases}
    		g\epsilon_c + g \Delta_c \lambda_c + g^2 \lambda_m = 0\\
    		\Delta_c \epsilon_m + g \Delta_c \lambda_c + \Delta_c \Delta_m \lambda_m = 0\\
    	\end{cases},
    \end{equation}
    \begin{equation}
    	\begin{cases}
    		\Delta_c \epsilon_m - g\epsilon_c + \qty(\Delta_c \Delta_m - g^2) \lambda_m = 0\\
    		g \lambda_c = -\epsilon_m - \Delta_m \lambda_m\\
    	\end{cases}, \quad
    	\begin{cases}
    		\lambda_m = \frac{\Delta_c \epsilon_m - g\epsilon_c}{\Delta_c \Delta_m - g^2}\\
    		\lambda_c = -\frac{\epsilon_m}{g} - \frac{\Delta_m}{g} \frac{\Delta_c \epsilon_m - g\epsilon_c}{\Delta_c \Delta_m - g^2}\\
    	\end{cases}.
    \end{equation}
    The resulting Hamiltonian is
    \begin{align}
    	H' 
    	&= H_0 + \frac 1 2 \qty[\Lambda, V] + \frac{1}{2} \qty[\Lambda, \qty[\Lambda, V]] + \mathcal O (\Lambda^3)\\
    	&=
    	\hbar \Delta_c c^\dagger c + \hbar \Delta_m m^\dagger m + \hbar g\qty(cm^\dagger + c^\dagger m) \\
    	&\quad
    	+ \frac{1}{2} \qty[\lambda_c c- \lambda_c^* c^\dagger, \hbar\qty(\epsilon_c c +  \epsilon_c^* c^\dagger)] 
    	+ \frac{1}{2} \qty[\lambda_m m- \lambda_m^* m^\dagger, \hbar\qty(\epsilon_m m +  \epsilon_m^* m^\dagger)] \nonumber\\
    	&=
    	\hbar \Delta_c c^\dagger c + \hbar \Delta_m m^\dagger m + \hbar g\qty(cm^\dagger + c^\dagger m)
    	+ \frac{\hbar}{2} \qty(\lambda_c \epsilon_c^* + \lambda_c^* \epsilon_c + \lambda_m \epsilon_m^* + \lambda_m^* \epsilon_m) \nonumber.
    \end{align}
    Note that the transformation is exact, since higher order commutators vanish (indeed, this is simply a displacement transformation). Furthermore, the last term is simply a constant energy offset with not physical effect, so it can be discarded.
    
    The transformed Hamiltonian $H'$ tells us that under coherent driving of both photons and magnons, the spectrum is equivalent to that of the undriven Hamiltonian $H$ after the replacements $(\omega_c,\omega_m \mapsto (\widetilde \omega_c - \omega_d, \widetilde \omega_m-\omega_d)$, with $\omega_d/2\pi$ the frequency of the drive. Hence, the standard level repulsion is obtained, and no level attraction can occur.


    \section{\label{app:comsol}Finite element modelling results}
\begin{figure}[t]
    	\centering
    	\includegraphics[width=0.5\linewidth]{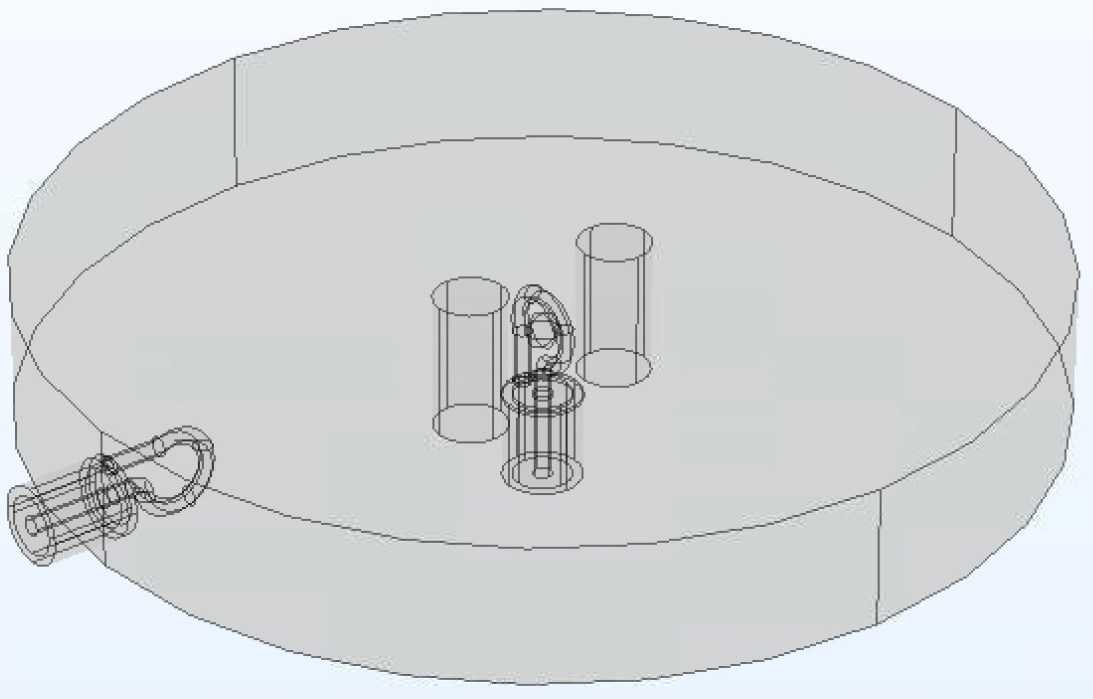}
    	\includegraphics[width=0.4\linewidth]{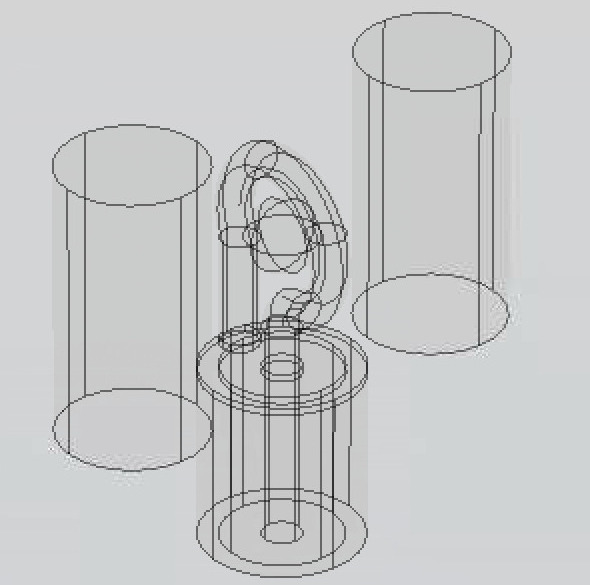}
    	\caption{Model of the two-post cavity we used in the \comsol\ simulations. The probes for Port 1 and Port 3 are an exact copy (dimensions and materials) limiting the effects of potential dephasings $\phi_1$ and $\phi_3$.}
    	\label{fig:comsol-model}
    \end{figure}

    \paragraph{Cavity design.}
    In this note, we detail the numerical results obtained using the RF module \comsol\ to simulate the two-tone driving experiment described by figure 1 in the main text. In the main text, $p_1^\text{in}$ and $p_3^\text{in}$ are defined as the signals output by the vector network analyser. Due to differing cable lengths or the geometry of the microwave probes, these signals can be dephased modelled using the phases $\phi_1$ and $\phi_3$. To limit these effects as much as possible, we chose to use identical probes for Port 1 and Port 3. To that effect,the probe for Port 3 is inserted from the bottom of the cavity instead of on the side, as pictured in \cref{fig:comsol-model}.
    
\begin{figure}[t]
    	\centering
    	\includegraphics[width=\linewidth]{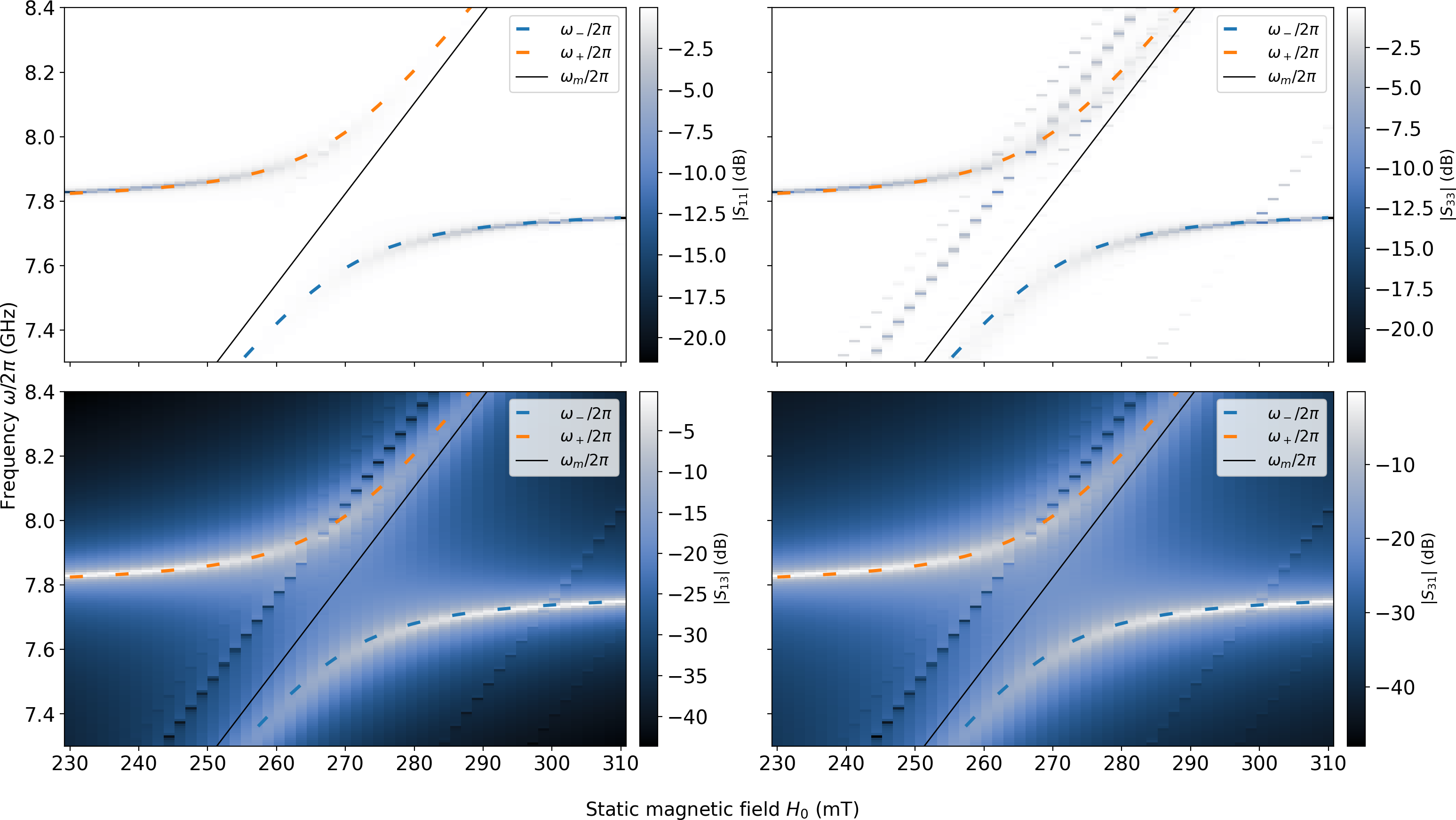}
    	\vspace{-2em}
    	\caption{Numerical calculations of the amplitude of the $S$ parameters of the cavity. The polaritonic frequencies $\omega_\pm/2\pi$ are given by equation (8) of the main text, and describe level repulsion, the signature of coherent coupling physics. The fit parameters are a coupling strength of $g/2\pi = 210$ MHz and a cavity resonance at $\omega_c/2\pi=7.853$ GHz.}
    	\label{fig:comsol-s-params}
    \end{figure}

    \paragraph{S parameters.}
    We first performed a frequency domain simulation to obtain the S-parameters of the cavity, shown in \cref{fig:comsol-s-params}. As expected, we observe energy level repulsion due to the coherent coupling of the photon and magnon in both reflection and transmission. Therefore, these simulations show that the physics of this system match with that predicted by the Hamiltonian model of equation (1) of the main text. We also note the presence of higher-order magnon modes corresponding to diagonal lines offset from the ferromagnetic frequency $\omega_m/2\pi$. This is especially the case for $S_{33}$, potentially due to the non-uniform magnetic field generated by the loop antenna.

\begin{figure}[H]
    	\centering
    	\includegraphics[width=\linewidth]{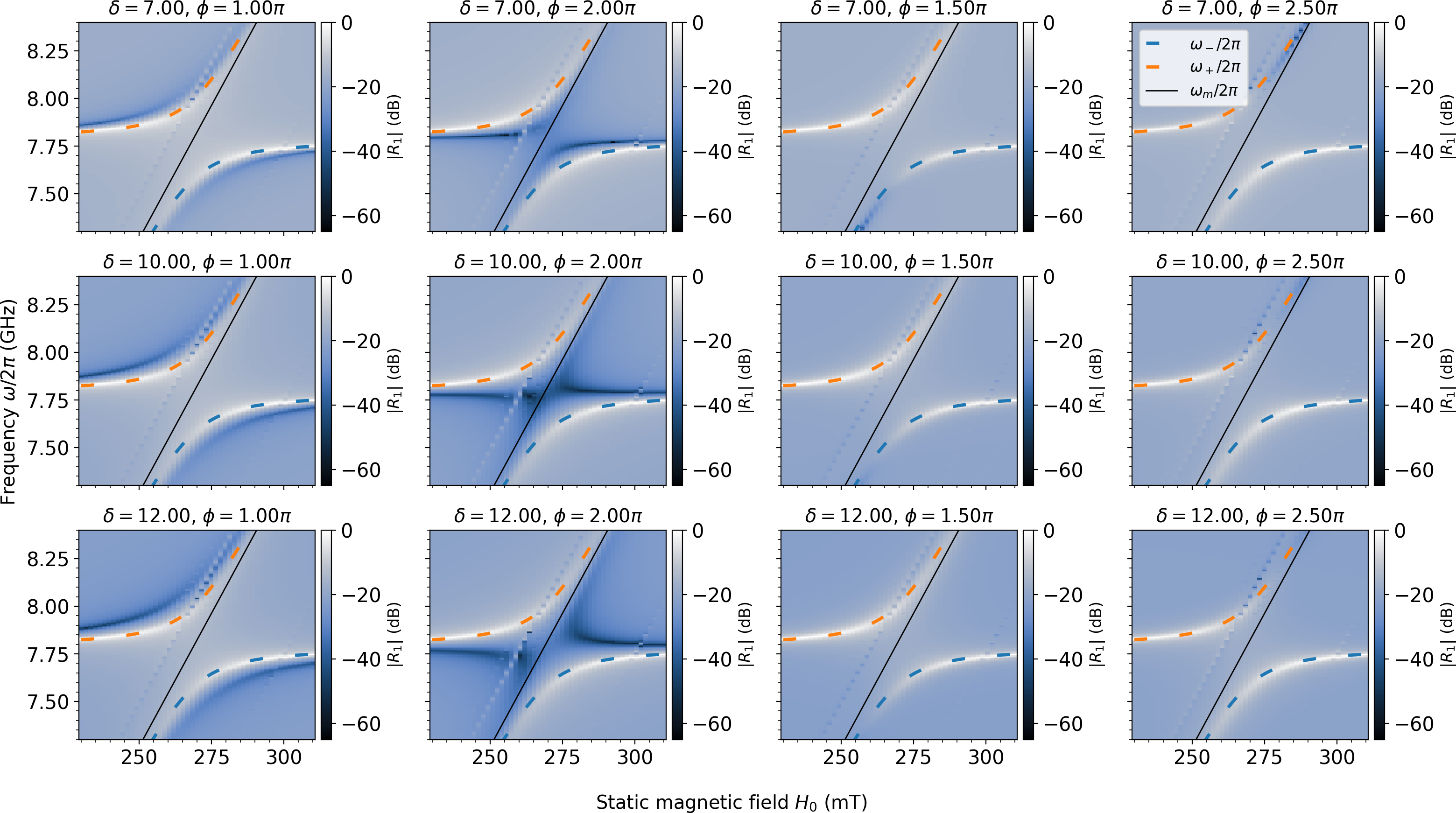}
    	\vspace{-2em}
    	\caption{Numerical calculations of the reflection $R_1$ using the S parameters obtained using \comsol. The legend is common to all figures The fit parameters are identical to this in \cref{fig:comsol-s-params}.}
    	\label{fig:prediction-R1-from-comsol-s-params}
    \end{figure}

    \paragraph{Estimation of $\delta$ and $\phi_0$.}
    Next, we used the numerical values of $S_{11}$ and $S_{13}$ to compute and plot $R_1 = S_{11} + \delta e^{i\phi}S_{13}$ for a range of $\delta$ and $\phi$. The results, plotted in \cref{fig:prediction-R1-from-comsol-s-params}, allow us to estimate very simply numerically which value of $\delta$  and $\phi$ are required to obtain level repulsion or attraction. We see that level repulsion and attraction are clearly visible for $\delta=12$ and $\phi \in \qty{\pi, 0}$ as predicted by the theory. This implies that the phase offset $\phi_0 = \phi_1-\phi_3 + \arg{g}-\frac{\pi}{2}$ in this system is $\phi_0=\pi$ since level repulsion is obtained when $\phi-\phi_0 = \pi-\phi_0 = 0$. Alternatively, one can extract the linewidths $\kappa_1/2\pi,\kappa_3/2\pi$ and the coherent coupling $\abs{g}/2\pi$ from the S-parameters of \cref{fig:comsol-s-params}, and then use $\delta_0 =\frac{\sqrt{\kappa_1 \kappa_3}}{\abs{g}} \delta$ to estimate $\delta$. Then, one needs to sweep the dephasing $\phi$ between the two drives to determine the phase offset $\phi_0$.
    
    \begin{figure}[H]
	    \centering
	    \includegraphics[width=\linewidth]{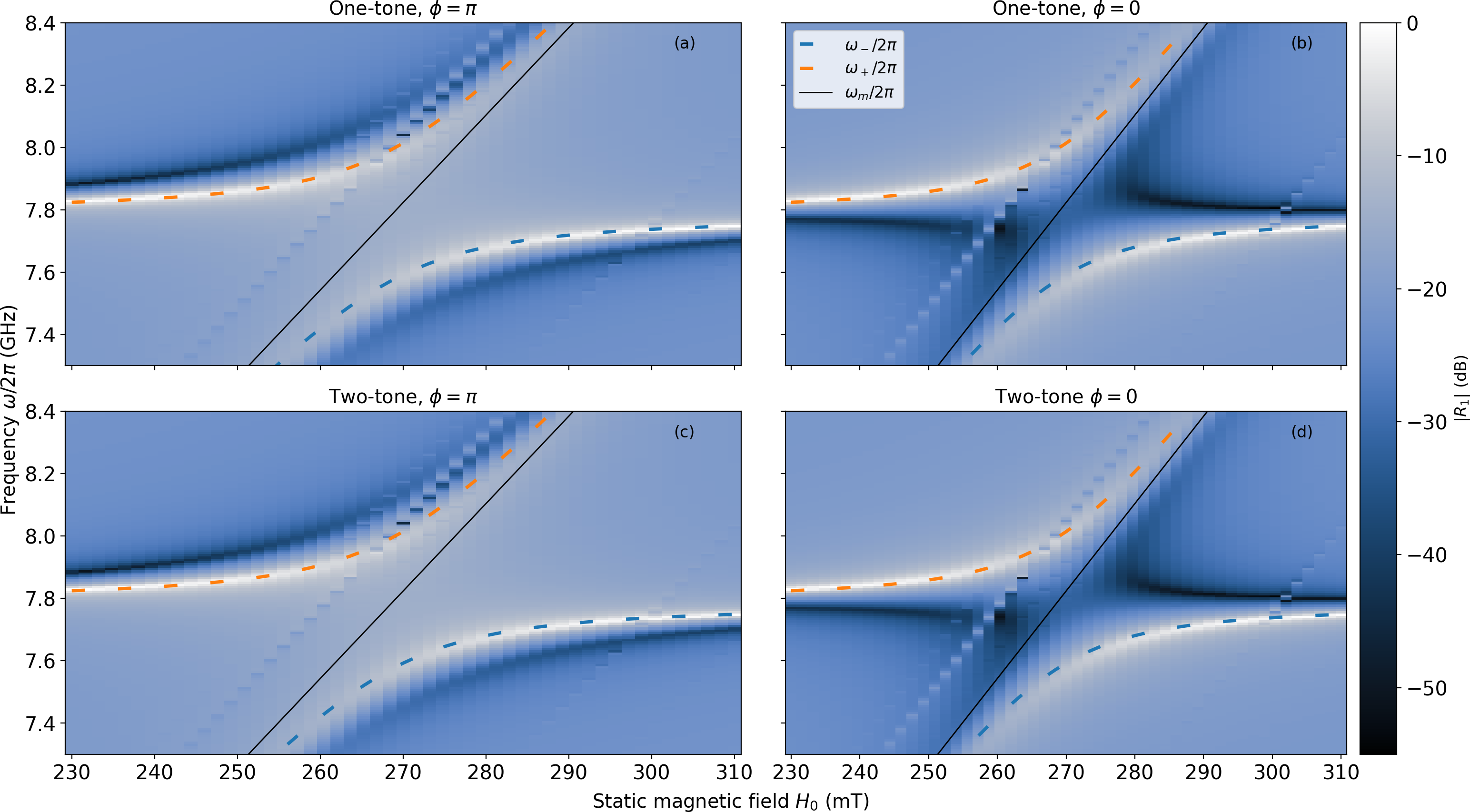}
	    \vspace{-3em}
	    \phantomsubfloat{\label{fig:comsol-two-tone:a}}
	    \phantomsubfloat{\label{fig:comsol-two-tone:b}}
	    \phantomsubfloat{\label{fig:comsol-two-tone:c}}
	    \phantomsubfloat{\label{fig:comsol-two-tone:d}}
	    \caption{Numerical calculations of the reflection at Port 1 with Port 3 active ($R_1$) using \comsol\ for $\delta=12$. (a-b) Evaluation of $R_1$ from the S parameters using $R_1 = S_{11} + \delta e^{i\phi}S_{13}$. By definition of the S parameters, these are essentially a combination of one-tone simulations. (c-d) Fully two-tone simulation where both Port 1 and Port 3 are simultaneously active.}
	    \label{fig:comsol-two-tone}
    \end{figure}

    Having determined a suitable $\delta$, we can now perform a true two-tone driving experiment by enabling both Port 1 and Port 3 in \comsol. The results for $\delta=12$ (\cref{fig:comsol-two-tone:c,fig:comsol-two-tone:d}) are compared with the manual evaluation of $R_1$ using the S parameters (\cref{fig:comsol-two-tone:a,fig:comsol-two-tone:b}). We observe a perfect agreement between the two.

	\bibliography{two-tone-driving}
\end{document}